\RequirePackage{fix-cm}
\documentclass{svjour3}                     
\smartqed  
\usepackage{graphicx}

\usepackage{amssymb}  
\usepackage{amsmath}
\usepackage{bm}
\usepackage{subfigure}
\usepackage{tabularx} 
\usepackage{tabulary}
\usepackage{color}
\usepackage{float}
\usepackage{array}
\usepackage{wrapfig}
\usepackage{multirow}
\usepackage{booktabs}

\usepackage{soul}
\usepackage{siunitx}
\usepackage{comment}
\usepackage{natbib}
\title{Bibliography management: \texttt{natbib} package}

\usepackage{chngcntr}
\counterwithin{figure}{section}

\usepackage{lineno}

\usepackage{latexsym}
 \journalname{ArXiv}
\begin{document}

\title{The Mars Microphone onboard SuperCam}
\titlerunning{SuperCam Microphone}        

\author{David Mimoun \and Alexandre Cadu \and Naomi Murdoch  \and Baptiste Chide    \and Anthony Sournac \and Yann Parot \and Pernelle Bernardi  \and  P. Pilleri \and Alexander Stott \and Martin Gillier \and Vishnu Sridhar \and Sylvestre Maurice \and Roger Wiens  \and the SuperCam  team}

\institute{D. Mimoun \at
              Institut Sup\'{e}rieur de l'A\'{e}ronautique et de l'Espace (ISAE-SUPAERO), Universit\'{e} de Toulouse, 31055 Toulouse Cedex 4, France \\
              \email{david.mimoun@isae.fr}           
            \\
            A. Cadu \at
              Institut Sup\'{e}rieur de l'A\'{e}ronautique et de l'Espace (ISAE-SUPAERO), Universit\'{e} de Toulouse, 31055 Toulouse Cedex 4, France \\
              \email{alexandre.cadu@isae.fr}           
            \\
            N. Murdoch \at
              Institut Sup\'{e}rieur de l'A\'{e}ronautique et de l'Espace (ISAE-SUPAERO), Universit\'{e} de Toulouse, 31055 Toulouse Cedex 4, France \\
              \email{naomi.murdoch@isae.fr}           
            \\
            B. Chide \at
              Institut de Recherche En Astrophysique et Planétologie, Toulouse, France \\
              \email{baptiste.chide@irap.omp.fr}           
            \\
            A. Sournac \at
              Institut Sup\'{e}rieur de l'A\'{e}ronautique et de l'Espace (ISAE-SUPAERO), Universit\'{e} de Toulouse, 31055 Toulouse Cedex 4, France \\
              \email{Anthony.Sournac@isae.fr}           
              \\
            Y. Parot \at
              Institut de Recherche En Astrophysique et Planétologie, Toulouse, France \\
              \email{yann.parot@irap.omp.fr}           
            \\
            P. Bernardi \at
              Laboratoire d'Etudes Spatiale et d'Instrumentation en Astrophysique (LESIA), Paris, France 
              \email{pernelle.bernardi@obspm.fr}           
             \\
            P. Pilleri \at
              Institut de Recherche En Astrophysique et Planétologie, Toulouse, France \\
              \email{paolo.pilleri@irap.omp.fr}           
            \\
            A. Stott \at
              Institut Sup\'{e}rieur de l'A\'{e}ronautique et de l'Espace (ISAE-SUPAERO), Universit\'{e} de Toulouse, 31055 Toulouse Cedex 4, France \\
              \email{Alexander.Stott@isae.fr}           
               \\  
            M. Gillier \at
              Institut Sup\'{e}rieur de l'A\'{e}ronautique et de l'Espace (ISAE-SUPAERO), Universit\'{e} de Toulouse, 31055 Toulouse Cedex 4, France \\
              \email{Martin.Gillier@isae.fr}           
              \\  
            V.  Sridhar \at
              Jet Propulsion Laboratory, 4800 Oak Grove Dr, Pasadena, CA 91109, United States \\
              \email{vishnu.sridhar@jpl.nasa.gov}           
              \and
           S. Maurice \at Institut de Recherche En Astrophysique et Planétologie, Toulouse, France \\
           \email{sylvestre.maurice@irap.omp.eu}
            \and
           R. C. Wiens \at 
           Los Alamos National Laboratories, NM 87544, USA,  \\
               \email{rwiens@lanl.gov}
           }

\date{January 2022}
\maketitle

\begin{abstract}
The “Mars Microphone” is one of the five measurement techniques of SuperCam, an improved version of the ChemCam instrument that has been functioning aboard the Curiosity rover for several years. SuperCam is located on the Rover’s Mast Unit, to take advantage of the unique pointing capabilities of the rover’s head. In addition to being the first instrument to record sounds on Mars, the SuperCam Microphone can address several original scientific objectives: the study of sound associated with laser impacts on Martian rocks to better understand their mechanical properties, the improvement of our knowledge of atmospheric phenomena at the surface of Mars: atmospheric turbulence, convective vortices, dust lifting processes and wind interactions with the rover itself.  The microphone will also help our understanding of the sound signature of the different movements of the rover: operations of the robotic arm and the mast, driving on the rough floor of Mars, monitoring of the pumps, etc ... The SuperCam Microphone was delivered to the SuperCam team in early 2019 and integrated at the Jet Propulsion Laboratory (JPL, Pasadena, CA) with the complete SuperCam instrument. The Mars 2020 Mission launched in July 2020 and landed on Mars on February 18th 2021. The mission operations are expected to last until at least August 2023. The microphone is operating perfectly.

\end{abstract}
\maketitle

\section{The Mars Microphone}

In July 2020, NASA launched a Rover that landed on Mars in February 2021 and has been operating since then on the surface of Mars, in the remnants of the Jezero crater delta which may include ancient sedimentary deposits and water altered materials \citep{mangold2021perseverance}. The Mars 2020 rover, named Perseverance, is dedicated to the study of Mars Habitability and the search of potential traces of ancient life, and to the study of the capacity of the explored locations environment to sustain life or the potential that had the visited sites to support life \citep{farley2020mars}. Like in the previous Mars Science Laboratory (MSL)-“Curiosity” Mars rover, the mission includes a long-duration science laboratory. Mars 2020 “Perseverance” rover is capable enough to make in-situ, multi criteria evaluation of samples, and to encapsulate them in “sample-return” containers left behind for future sample return missions \citep{muirhead2020mars}. Of course, the assessment of present and past habitability includes multidisciplinary measurements; habitability criteria require a thorough evaluation in various thematic fields, such as biology, climatology, mineralogy, geology and geochemistry.
Among the scientific instruments on board “Perseverance”, SuperCam \citep{Wiens2021} provides a rich set of tools to help the M2020 “Perseverance” rover to reach those scientific goals.
The SuperCam instrument is an evolution from the successful ChemCam instrument on MSL- “Curiosity” \citep{maurice2012chemcam}. SuperCam is an instrument package capable of four different remote-sensing techniques: Laser-Induced Breakdown Spectroscopy (LIBS), Raman and time-resolved fluorescence (TRF), passive visible and infrared (VISIR) reflectance spectroscopy, and remote micro-imagery (RMI). A fifth technique, the sound recording, has been added to complement the LIBS measurements and also to open a new window of measurements on Mars: prior to Mars 2020, no sounds had ever been recorded on the surface of Mars.

\subsection{A brief history of planetary microphones}

The SuperCam Microphone is not by far the first microphone that has been implemented in a space mission, but it has been the first to operate successfully on Mars and to record the first sounds of Mars. The idea of having sounds from Mars, and more generally from other worlds, has an incredible popularity among the general public. The short history of the planetary microphones probably began with the Grozo 2 instrument during the Venus Venera 13 and 14 missions \citep{ksanfomaliti1982acoustic}. 

A successful attempt to record the `sound' of the Huygens probe entry, descent and landing on Titan was made in 2005. Even though this `sound' was actually reconstructed from accelerometer data recorded by the Huygens Atmospheric Structure Instrument during its descent through the atmosphere of Titan, it has still been a popular success, downloaded thousands times from the European Space Agency (ESA) website, see e.g. \citep{leighton2004sound}.


The original Mars Microphone instrument, funded by the Planetary Society \citep{delory2007development} was built for the ill-fated NASA Mars Polar Lander (MPL) mission, which lost contact with Earth shortly after its descent to the Martian surface and was never recovered. However, the worldwide interest in the Mars Microphone project was so intense that immediately following the loss of MPL, an opportunity to fly the microphone experiment was provided by the Centre National d’Etudes Spatiales (CNES), the French Space Agency, on the NetLander mission \citep{dehant2004network} to Mars in 2007. But NetLander was cancelled in 2001, and the Mars microphone was, therefore once again relocated, now to operate on the Phoenix MARDI camera. On the Phoenix Mission \citep{smith2004phoenix} the same team provided the microphone. Unfortunately the MARDI Camera was not operated during the Phoenix mission, for fear of major electrical interference with a high-priority instrument.

Finally, a last proposal was made by ISAE Team to ESA for the (also ill-fated) Schiaparelli descent module. The microphone was planned to be integrated into the DREAMS payload package \citep{esposito2013dreams} as an add-on to other atmospheric science payloads. The microphone goals were to detect, during the short life of the lander on the Mars ground (up to three days in the most optimistic case), atmospheric related events such as convective vortices, sand saltation noise or any other atmospheric noise. However, some reserves were issued by ESA on the possible science outcome of the instrument and it was finally not implemented.

\subsection{SuperCam instrument overview}
\label{Supercam-description}

The SuperCam instrument is an evolution from the successful ChemCam instrument on MSL-Curiosity\citep{maurice2012chemcam}. In addition to the geologic investigation capabilities, linked  to the Laser Induced Breakdown Spectroscopy, or LIBS technique (see Section \ref{LIBS-science}) it implements a new Raman biologic spectroscopic analysis, which is coupled to an Infra-Red (IR) spectrometer.  Another improvement has been made by the addition of colour to  Remote Micro Imager (RMI), which provides context for the instrument.

\begin{figure}[ht]
\noindent\includegraphics[width=0.9\textwidth]{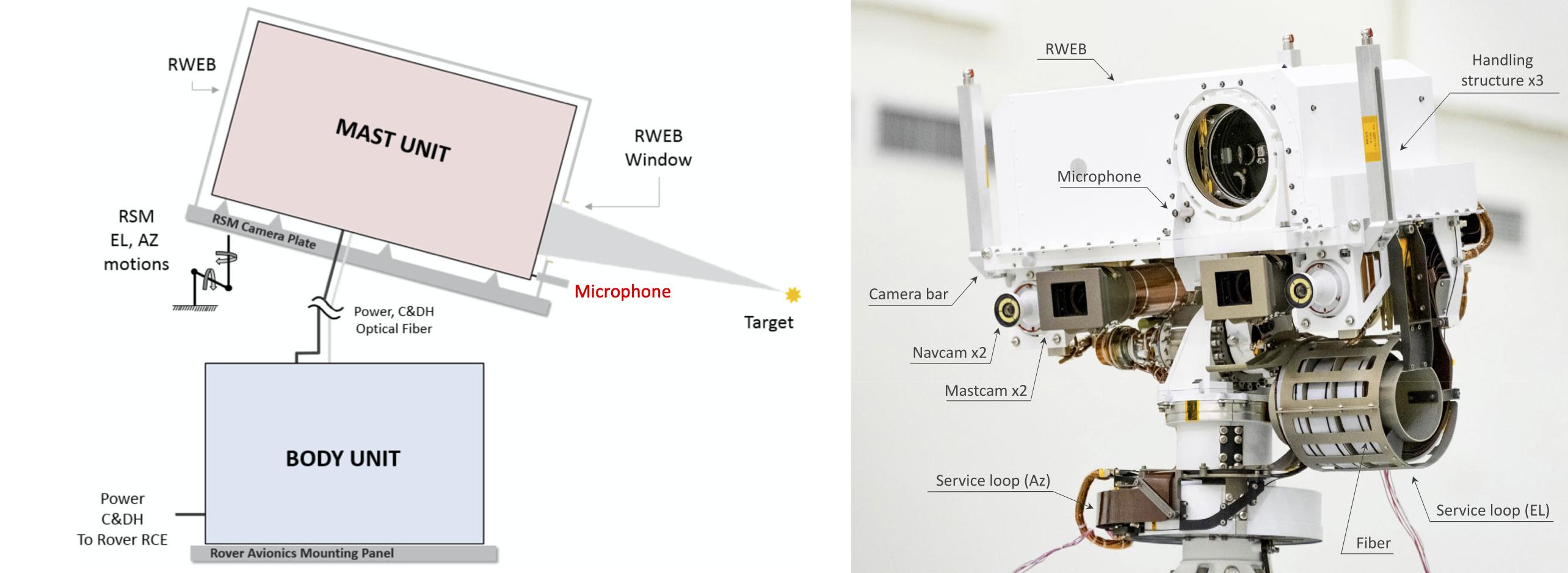}
\caption{SuperCam Block diagram (left) - figure modified from \citep{Maurice2021} SuperCam Mast Unit (right) after its integration at JPL - Credits NASA/JPL }
\label{fig:SuperCam}
\end{figure}

The SuperCam package consists of three separate major units: the “Body Unit”, the “Mast Unit” and the “Calibration Targets” (see figure \ref{fig:SuperCam}) .
The Mast Unit (MU) consists of a telescope with a focusing stage, a pulsed laser and its associated electronics, an infrared spectrometer, a color CMOS micro-imager, and focusing capabilities. A new development for SuperCam is separate optical paths for LIBS (“red line”) and Raman spectroscopy (“green line”), which produces a frequency-doubled beam.

The Body Unit (BU) consists of three spectrometers covering the UV, violet, and visible and near-infrared ranges needed for LIBS. The UV and violet spectrometers are identical to ChemCam. The visible spectrometer uses a transmission grating and an intensifier so that it can double as the Raman spectrometer. The intensifier allows the rapid time gating needed to remove the background light so that the weak Raman emission signals can be easily observed. A fiber optic cable, as well as signal and power cable, connects the Mast and Body units.

In addition, a set of calibration targets (CT) mounted on the Rover will enable periodic calibration of the instrument. A complete description of the instrument can be found in \citep{Maurice2021}.

The Mast Unit is provided by IRAP (funding from CNES), while Los Alamos National Laboratories (LANL, US) provided the Body Unit. The IRAP and LANL portions are entirely separate mechanically, greatly simplifying the interface controls as well as development across international boundaries. The University of Valladolid (UVa) in Spain is the lead for the SuperCam on-board Calibration Target .

\section{SuperCam Microphone Science Objectives}

\subsection{Science Objectives derived from SuperCam}

The SuperCam Microphone goal is to record audio signals on the surface of Mars from both natural and artificial origins. Contrarily to previous attempts to operate a microphone on Mars, which were primarily for outreach purposes, the primary science objective is to support the SuperCam LIBS investigation. LIBS stands for "Laser Induced Breakdown Spectroscopy" and is the key technique that has been developed in the frame of the ChemCam experiment \citep{maurice2012chemcam} to analyse - at distance - the composition of the martian rocks. It uses a powerful laser to ablate rocks and create a plasma: the emitted radiation is then be collected by a telescope and its spectrum is analysed (see e.g. figure \ref{fig:LIBS-sound}).  

The sound recording of the LIBS laser shots provides a unique opportunity to obtain the properties of Martian rocks and soils, mostly related to the rock hardness \citep{Maurice2021, Chide2019, Murdoch2019}. This objective is directly linked to the primary science of SuperCam, however, the SuperCam microphone also provides several other scientific opportunities. By providing wind and turbulence measurements, and potentially recording dust devils at a close distance \citep{Chide2021,murdoch2021predicting,murdoch2021ATM}, the SuperCam Microphone will also contribute to the Mars 2020 atmospheric science goals linked to the circulation, weather and climate, the dust cycle and even aeolian processes. From an engineering perspective, the SuperCam microphone can also enable backup determination of the SuperCam telescope focus \citep{lanza2021expected}, and monitoring of artificial sounds emitted by other payloads (e.g. MOXIE \citep{hecht2021mars}, or Z-CAM \citep{bell2021mars}), or by the rover operation itself. 
If we refer to SuperCam Science objectives and goals \citep{Maurice2021}, the SuperCam Microphone complements the experiment for the following goals:

\begin{figure}[ht]
\noindent\includegraphics[width=0.95\textwidth]{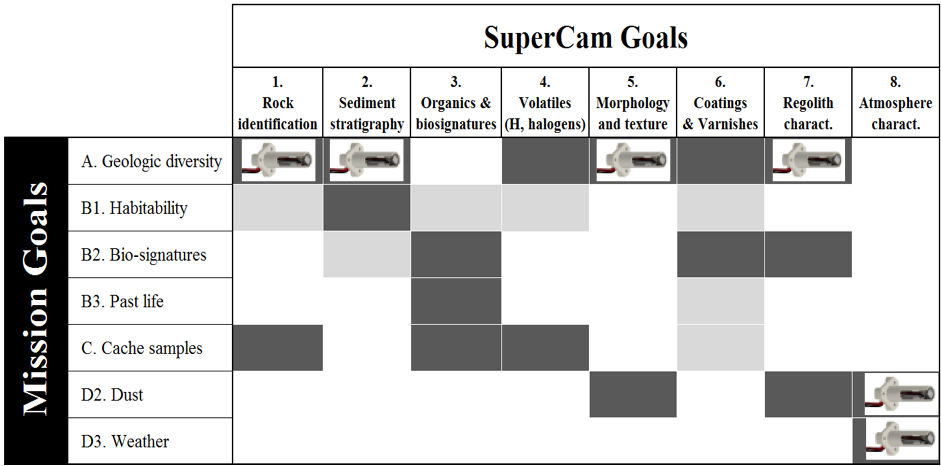}
\caption{SuperCam science goals as described in \citep{Maurice2021}.The microphone pictures indicate the goals to which the microphone contributes}
\end{figure}
\begin{figure}[ht]
\begin{tabular}{ | m{7.5em} | m{4cm}| m{4cm} | } 
  \hline
    & \textbf{Objectives} & \textbf{Comments} \\ 
  \hline
  \textbf{SuperCam Goal 1/2/5} & Rock Identification, sediments morphology and texture & Characterization of target properties:sound wave amplitude depends on ablated material quantity at a given distance, e.g. \citep{GRAD1993370} \\ 
  \hline
  \textbf{SuperCam  Goal 7}  & Regolith Characterization & SuperCam microphone will help addressing soil properties: the acoustic energy of the sound wave depends on material compaction and  hardness  \citep{QIN2004325,Chide2019,Murdoch2019} \\ 
  \hline
  \textbf{SuperCam Goal 8}  & Atmospheric Characterization  & SuperCam microphone secondary goals will help addressing various atmospheric phenomena such as wind properties, dust devils, turbulence, etc. \citep{Chide2021,murdoch2021predicting,murdoch2021ATM} \\ 
  \hline
\end{tabular}

\caption{Description of the Microphone goals}
\label{fig:SuperCamGoals}
\end{figure}

Finally, it is important to recall that this instrument opens a new window in our Mars observation capabilities, adding for the first time the sense of "hearing" to a rover. The SuperCam Microphone is, therefore, a powerful outreach tool, which draws a lot of general public attention to planetary science.
\subsection{Sound propagation in the Martian Atmosphere}
\label{sound_propagation}
Among the various reasons given by mission or review boards not to support the selection of microphones on missions to Mars, one comes back very often: many scientists believe that a Mars microphone would record hardly anything, due to the combination of low pressure of the atmosphere and of the expected attenuation of the sound in a carbon dioxide atmosphere.

\begin{figure}[ht]
\noindent\includegraphics[width=0.9\textwidth]{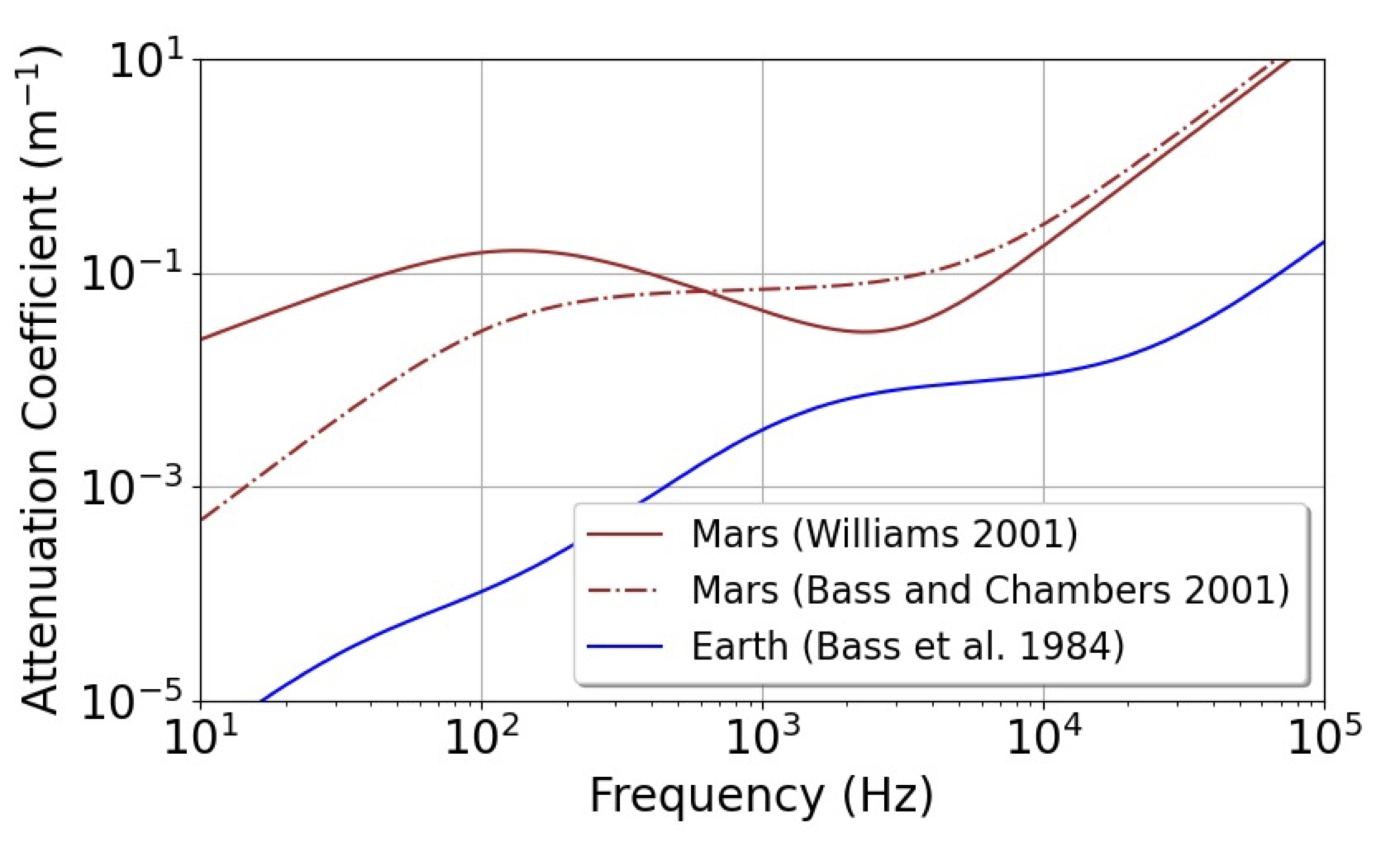}
\caption{Sound attenuation coefficient for an atmosphere of CO2 at 240K and 740 Pa. Based on the models from \citep{williams2001} (dashed red line), and \citep{bass2001absorption} (solid red line). A peak in absorption is predicted around 3 kHz. Also shown (solid blue line) is the frequency dependent attenuation coefficient for the Earth.}
\label{fig:attenuation_models}
\end{figure}

As a matter of fact, pioneering work had been done notably by \citep{williams2001} or \citep{bass2001absorption} for the Mars Polar Lander mission: sound propagation on Mars is expected to be similar to the Earth stratosphere, with an average atmospheric pressure between 6 and 8 millibars and a mean temperature of about 240 K.  These acoustic models predict a frequency-dependent sound speed and an attenuation where inflection points at a few kHz result from carbon dioxide molecular relaxation processes (See figure \ref{fig:attenuation_models}): in the cold, carbon dioxide Martian atmosphere, they predict a strong attenuation across the audible frequency range. We have used this model and considered that the acoustic pressure amplitude follows Equation \ref{equ:attenuation_model} 

\begin{equation}
  p(r,f)=p_0\left(\frac{r_0}{r} \right)^\beta .e^{-\alpha(f).r}
\label{equ:attenuation_model}
\end{equation}
where p is the pressure at a distance $r$, $r_0$ the reference distance, $p_0$ is the pressure at the source, and $f$ the frequency.  The $\beta$ coefficient derives from the geometric attenuation ($\beta=1$ for a spherical wave and $\beta=0$ for a plane wave front). $\alpha(f)$ is a frequency-dependent attenuation coefficient that is graphically represented in Figure  \ref{fig:attenuation_models}.

\vspace{2cm}
 
\begin{figure} 
\includegraphics[width=0.7\textwidth]{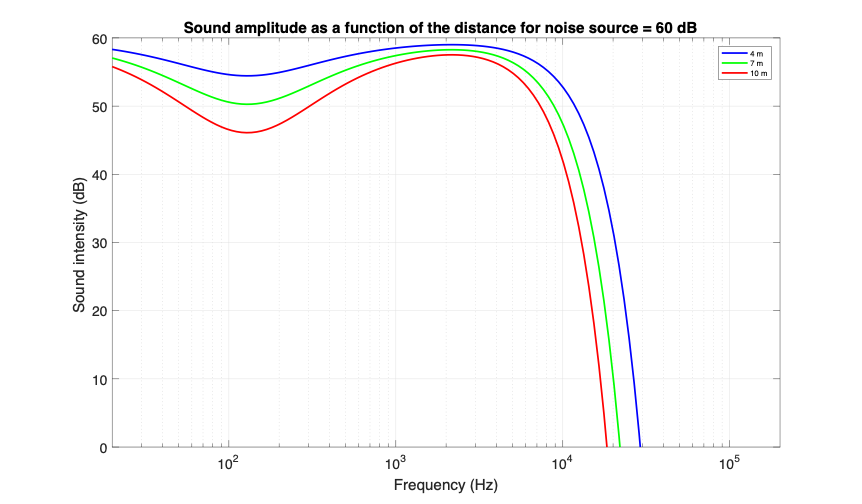}
\caption{The sound intensity of a 50 dB source as a function of frequency for three different distances (4 m - blue, 7 m - green and 16 m - red).   Derived using the model of \citep{williams2001}}
\label{fig:attenuation_sensitivity}
\end{figure}

 It is predicted that most sounds in the frequency range audible to the human ear (~20 Hz - 20 kHz) will not propagate over more than some tens of meters, particularly the higher frequency range (see \ref{fig:attenuation_sensitivity}). However, the situation improves in the lower frequencies and infra-sound region ($<$20 Hz). Such low acoustic frequencies, produced, for example, by dust devils (e.g.\citep{DustDevilInfrasoundSignatures}), or bolide impacts (e.g. \citep{williams2001}), could propagate over kilometer ranges. This process has been described e.g in \citep{Martire2020Infrasounds} in the context of the NASA Insight mission \citep{banerdt2016, banerdt2020}.
In order to assess the potential of sound recordings on Mars, and build a science focused Mars microphone, we have, therefore, chosen to use a double approach: design an instrument based on these analytical models, and confirm the performance of the instrument with tests in a Martian environment (see Section \ref{Test_Campaigns}). 

As it can be derived from figure \ref{fig:attenuation_sensitivity}, the sound is not expected to propagate significantly at frequencies over 10 kHz, and the maximum distance of recording of a sound of  typical  amplitude (e.g. 60 dB) is relatively limited. To this end, a nominal sampling frequency of 25kHz is chosen for the SuperCam microphone in order to capture most of the audible signal. An optional 100 kHz frequency has been added to allow the optimization of low pass filters. 

The start of each laser shot is easily detected though their electromagnetic impact on the recording. The difference between the time of this spike and the beginning of the sound arrival can be used to determine the average speed of sound along the targeted direction, providing that the distance between the SuperCam Mast Unit and the target is known. These distances can be known thanks to the 3D model of the environment built on the Z-Cam stereo view. This method for estimating the speed of sound is described in \citep{chide2020speed}.

\subsection{Recording the LIBS measurements}
\label{LIBS-science}

The first objective of the SuperCam Microphone is to record sounds resulting from the interaction of the laser with the rock targeted by the LIBS technique. 

\begin{figure}[ht]
\noindent\includegraphics[width=0.8\textwidth]{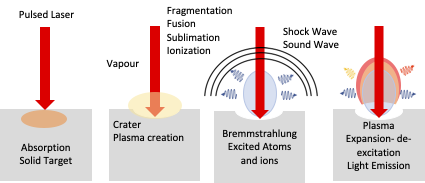}
\caption{LIBS : the process of sound emissions - after \citep{RehseLIBS} }
\label{fig:LIBS-sound}
\end{figure}

LIBS is a chemical analysis technology that uses a short laser pulse to create a micro-plasma on the surface of the sample. The micro-plasma is then analyzed by a spectrometer, which analyzes the spectrum of the sample to be studied. This technique was invented when the laser first appeared in the 1960s. The term LIBS was introduced in the 1980s in reference to the breakdown of the air by laser pulses during the creation of plasma. This technique, which allows the remote chemical analysis of samples (therefore without contact with potentially dangerous samples, such as radioactive samples), is now experiencing a renewed interest due to the appearance of powerful lasers delivering more powerful pulses. It is also a perfect tool for remote sensing when transporting samples to a scientific laboratory is challenging. The process by which the laser sparks creates a sound is described in Figure \ref{fig:LIBS-sound}.

\begin{figure}[ht]
\noindent\includegraphics[width=0.45\textwidth]{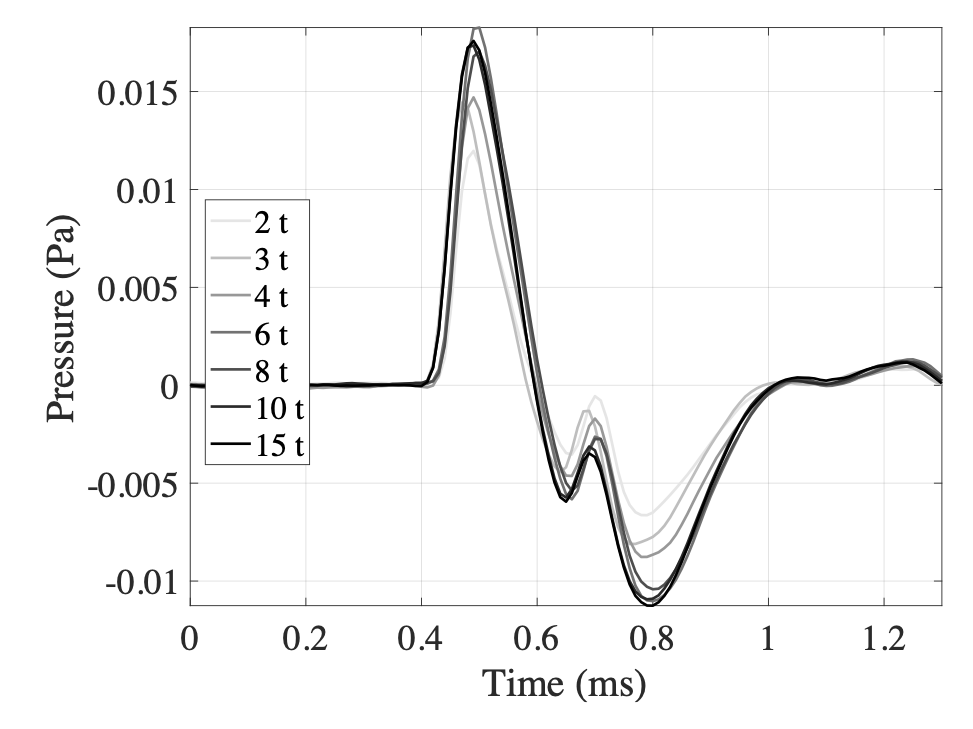}
\noindent\includegraphics[width=0.45\textwidth]{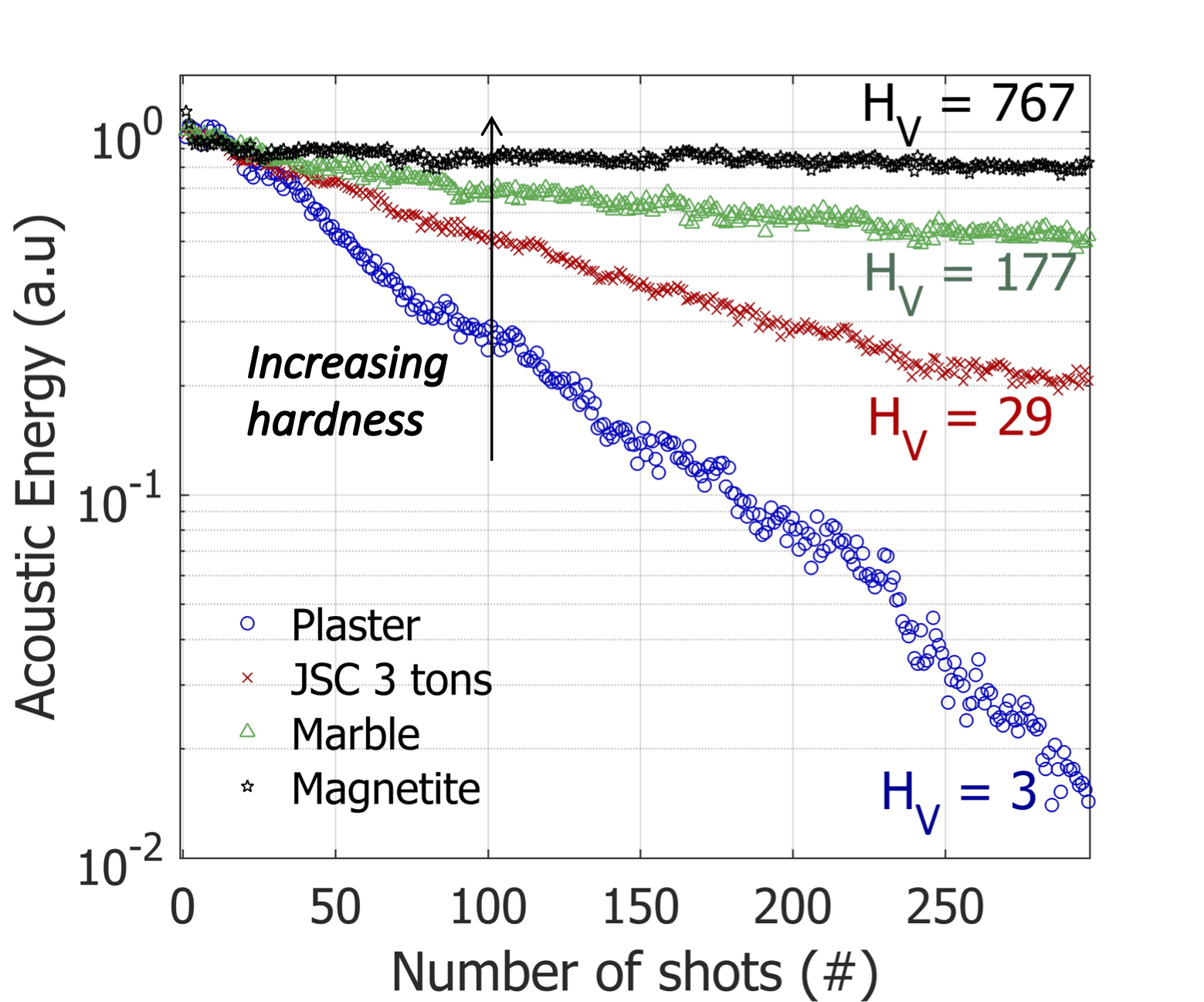}
\caption{Results of our experiments a - (Left) Waveform measured during impacts on samples of Martian soil simulants JSC1 at different levels of compaction - After \citep{Murdoch2019}. b- (Right) The rate of decay of acoustic energy during a series of shots depends on the hardness of the target. After \citep{Chide2019}.}
\label{fig:LIBS-hardness}
\end{figure}

The interest of this recording is two-fold: first of all, this technique allows remote analysis of the target, without bringing the robotic arm into contact with the mineral to be analyzed; In addition, the first laser impacts vaporize the layer of dust that covers the rock to be analyzed, allowing an analysis of the rock thus uncovered. However in the process the structure of the target is lost. Listening to LIBS sparks provides new information relative to the ablation process that is independent from the LIBS spectrum.  In the LIBS literature, the acoustic wave is known to be a product of laser-induced evaporation at high power density of radiation on a sample surface:

\begin{equation}
 \Delta P  \approx m_{abl}.1/v_{acw}. 1/r
\end{equation}

where $v_acw$ and $r$ are the velocity of acoustic wave front and the distance. Thus, the intensity of the acoustic signal acquired as the peak-to-peak amplitude of acoustic waveform will be proportional to the ablated masses $m_{abl}$ as demonstrated by \citep{chaleard1997correction} for aluminum alloys and \citep{GRAD1993370} who used various ceramics. In our experiments with the SuperCam microphone, \citep{Murdoch2019} used soil simulant targets, to demonstrate that the acoustic signal associated with the plasma formation during the LIBS experiment varied as a function of the target compaction. Then \citep{Chide2019} compared in detail the shot-to-shot evolution of acoustic energy with the laser induced crater morphology and plasma emission lines. The chosen targets are a set of geological targets of various origin and hardness; the depth and volume  of the craters created by the LIBS impacts have been profiled and analysed with the associated sound recording. The observable here is the acoustic energy recorded by the microphone. A good proxy of this acoustic energy is the integral of the waveform. The decrease of the acoustic energy as a function of the number of shots is well correlated with the target hardness/density (Figure \ref{fig:LIBS-hardness}). This can be explained as the acoustic energy source originates in the hole created by the sparks: shape and  volume of the hole vary as a function of the number of shots, changing the source geometric properties. 

Therefore, listening to laser-induced sparks can complement the LIBS experiment by providing constraints on the hardness/density of the targeted rock. An interesting consequence of the sensitivity of the acoustic signal to target hardness/density is that we can interpret a rupture of the shot-to-shot energy change as a proxy for the existence of a rock coating at the surface of the target, as described in \citep{lanza2020listening}.

\subsection{Atmospheric Science}

The key atmospheric science goal of the microphone is to characterize the Martian atmospheric dynamics at high frequency (much higher than Perseverance’s Mars Environmental Dynamics Analyzer - MEDA - instrument suite, which make measurement at up to 2 sample per second (sps) for the wind, and 1 sps for the pressure \citep{rodriguez2021mars}). 
The microphone measures high frequency variations in the dynamic pressure, and such pressure fluctuations are crucial to understanding the Martian climate, including the diurnal and seasonal evolution. Therefore, the atmospheric science investigations of the microphone, can be linked to the Mars 2020 mission high level atmospheric investigations and science goals:

\begin{itemize}
    \item What controls the circulation, weather and climate? 
    \item What controls the dust cycle? 
    \item Aeolian processes and rates 
\end{itemize}

\subsubsection{Measurement of the atmospheric turbulence}

Atmospheric turbulence is a key property of the Martian atmosphere. The thinness of its atmosphere associated to the thermal properties of its sandy surface pave the way for strong instabilities of the boundary layer gradient, resulting in convective turbulence (e.g. \citep{tillman1994boundary}). Pressure fluctuations and wind gusts (both observable on the microphone are manifestations of convective motions in the atmosphere. These convective motions are linked to dust motion and lifting, and there is also a link between the observed pressure fluctuations and the atmospheric opacity measurements (e.g., \citep{ullan2017analysis} )  

The turbulent properties of the Planetary Boundary Layer (PBL) are key to understanding the conditions of the Martian atmosphere (see \citep{spiga2018}, \citep{chatain2021seasonal}). Following our work on the Mars atmospheric turbulence spectrum \citep{murdoch2016,mimoun2017,temel2022,murdoch2022} based on wind, pressure and infra-sound measurements, we plan to use the SuperCam microphone to extend the measurements of the InSight and Perseverance meteorological suites to higher frequencies. The microphone will allow us to characterise the Martian dynamic pressure fluctuations at high frequency for the first time, giving us information about the spectral content of high frequency turbulence. 

In addition, the combination of MEDA and microphone data will allow us to investigate the full energy spectrum of the atmosphere, and its potential variations at hourly, daily and seasonal scales. In addition, we should be able to identify the transition frequency (or frequencies) between the various regimes of the Martian atmosphere at the Jezero site.

\subsubsection{Measurement of the wind speed}
\label{wind_speed}

\begin{figure}[ht]
\noindent\includegraphics[width=0.9\textwidth]{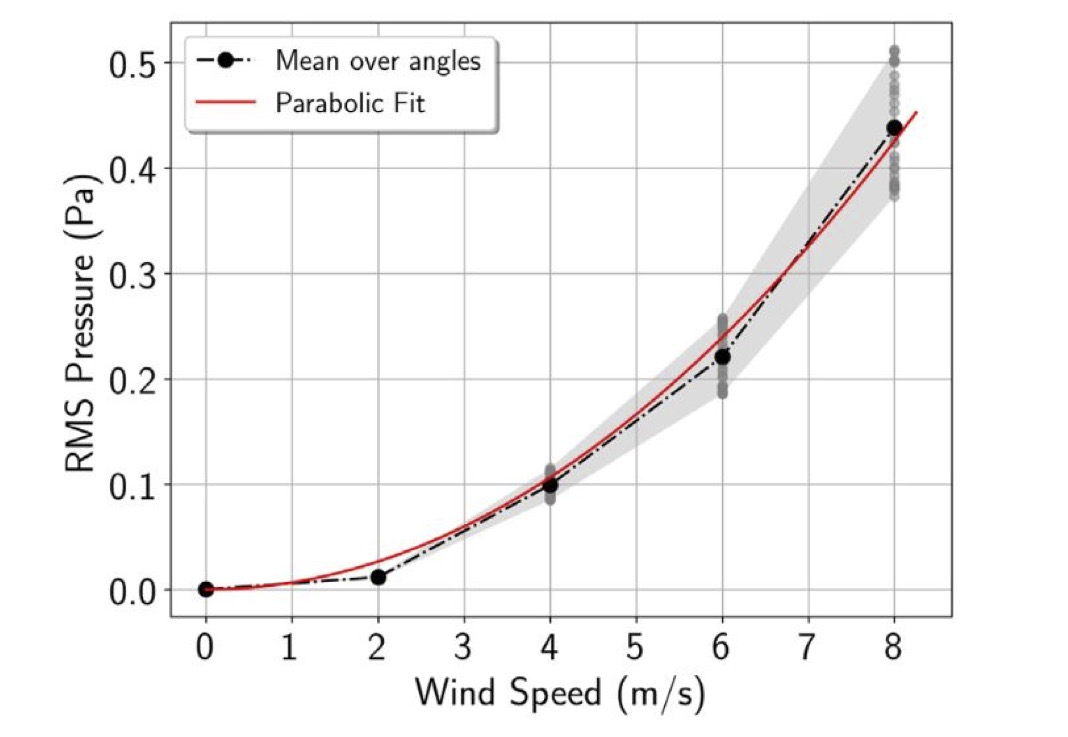}
\caption{Relationship between wind Speed and RMS sound pressure level measured by the microphone during Aarhus tests - After \citep{Chide2021}. RMS is calculated over the 100-500 Hz band. Gray dots are values for different angles of the Mast Unit wrt the wind.}

\label{fig:Wind_RMS}
\end{figure}

  A first attempt of measurement the wind speed thanks to an interplanetary microphone has been done on Venus \citep{ksanfomaliti1983wind}. This measurement is based on the the following relationship (as described in \citep{morgan1992investigation}.)\\
  
  \begin{equation}
                          P = \rho U V 
               \end{equation}
           
  \noindent where P is the sound pressure, $\rho$ the atmospheric density, $U$ the variation speed and $V$ the mean wind speed.  \citep{lorenz2017wind} use the Aarhus Martian wind tunnel to demonstrate the potential for using a microphone as a wind speed sensor on Mars. Specifically, they report that the Root Mean Square (RMS) voltage measured by the microphone varies with the wind speed. Similarly, during the end-to-end tests of the SuperCam microphone in the same wind tunnel (see Section \ref{Test_Campaigns} and \citep{Murdoch2019,Chide2021}), we have been able to correlate the wind speed to the microphone measurements (Figure \ref{fig:Wind_RMS}), and quantify the influence of the SuperCam mast-unit orientation with respect to the wind direction. 

As the wind tunnel is not fully representative of the environment on Mars, an in-situ calibration is required to be done after landing, with a 360° sampling of the sound measurement, while simultaneously measuring the wind speed and direction using MEDA.

\subsubsection{Acoustic detection of dust devils  and convective vortices:}

Dust devils are convective vortices, usually of a few meters to tens of meters in diameter, that  lift and transport dust particles.  This phenomenon has been witnessed on Earth since centuries in arid regions, or more generally in regions where the convective activity of the atmosphere is important \citep{Balme2006}. Vortices and dust devils are also common on Mars e.g. \citep{Ferri2003,Murphy2016,perrin2020monitoring}.

\begin{figure}[ht]
\noindent\includegraphics[width=0.45\textwidth]{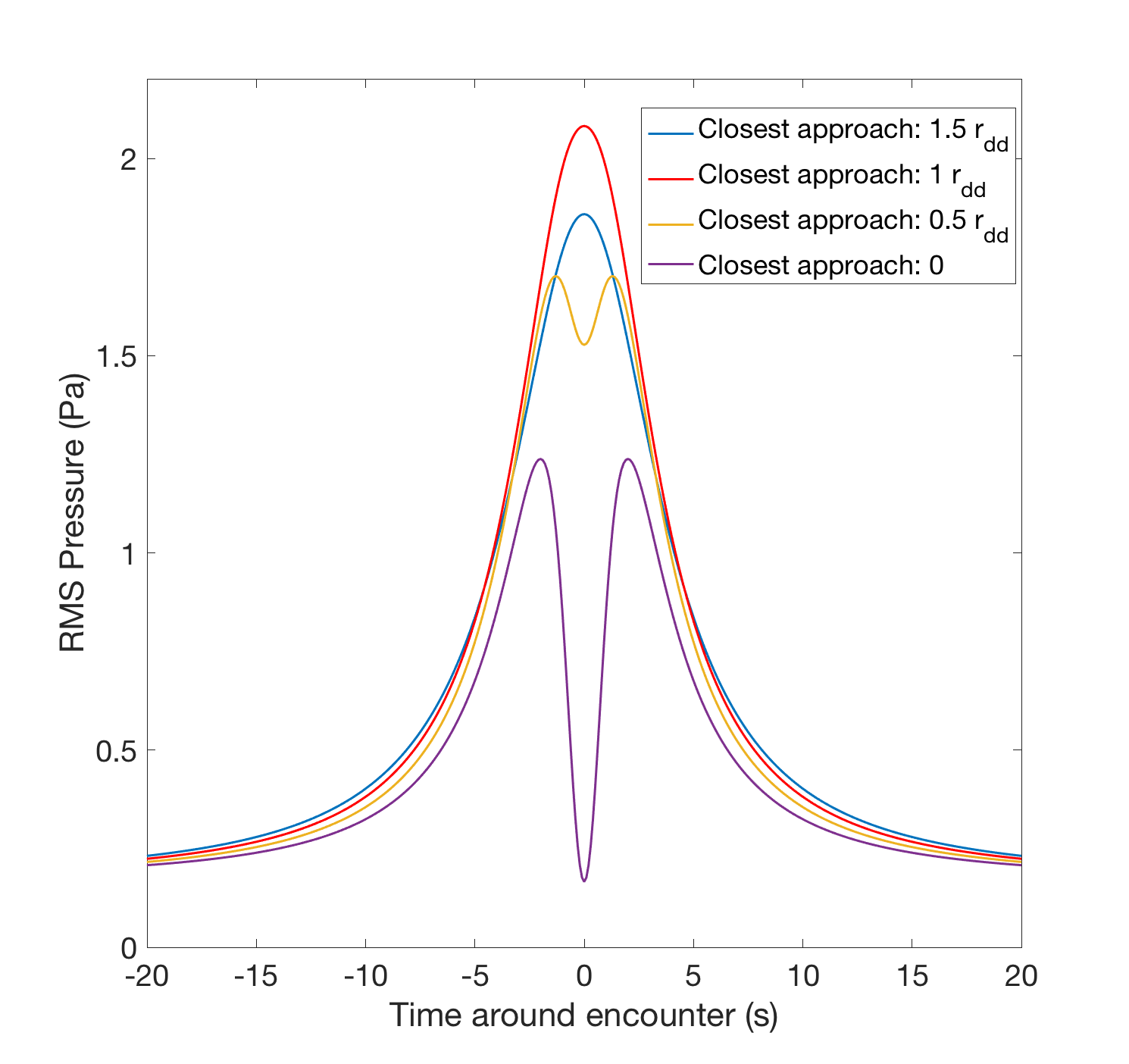}
\noindent\includegraphics[width=0.55\textwidth]{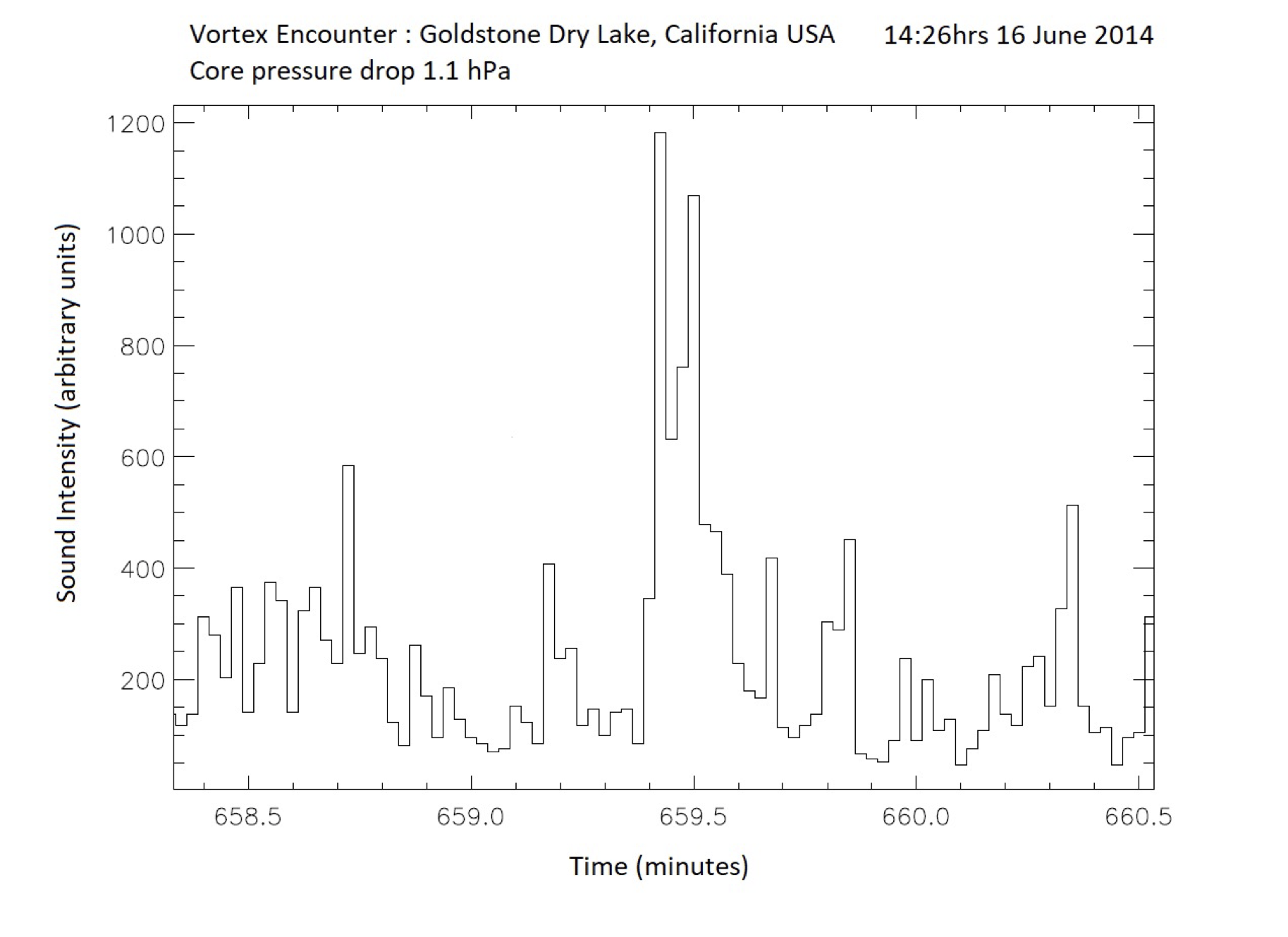}
\caption{(Left) Simulated SuperCam Microphone vortex signals as a function of the closest approach distance. Vortex parameters: 10 m radius, 5 Pa core pressure drop, advecting past the microphone at 5 m/s \citep{murdoch2021predicting}.  - (Right) Acoustic recording of a dust devil vortex \citep{murdoch2021predicting}}
\label{fig:dust_devils}
\end{figure}

As described in \citep{lorenz2015dust}, convective vortices have an infrasonic counterpart that has already been detected on Earth. We intend to follow up this research by attempting to record the sound pressure level associated with dust devil encounters. However, \citep{murdoch2021predicting} explain that the dominate vortex signal on the SuperCam Microphone will likely be the pressure fluctuations induced by the wind dynamic pressure (Figure \ref{fig:dust_devils}, left). Such a signal has also been observed with microphones in terrestrial field experiments \citep{lorenz2017wind, murdoch2021predicting}.  

In any case, in order to measure the vortex winds, or to record any possible infrasounds generated by vortices despite the sound attenuation of the Martian atmosphere, a very close-range vortex encounter will be necessary. This will probably need a dedicated measurement campaign (and some luck!).

\subsection{Rover and Other Sounds monitoring:}

The primary objectives of the SuperCam microphone are related to Mars 2020 science. However, these objectives can be completed by several opportunistic secondary objectives that can be classified in roughly two types of objectives: first,  "engineering support", with the recording of the various sounds produced by the Mars Perseverance Rover. As during our everyday life, listening to the noise from mechanical devices provides a quick diagnosis on the inner workings of a complex piece of engineering. The SuperCam microphone team has been in touch with various other instrument teams (e.g. MOXIE pumps \citep{hecht2021mars}), or MastCam to help them to record a "reference" noise recording that will serve to help investigate any later issue in the instrument operation. The table \ref{tab:various-noise} makes a summary of possible sound recordings.

Second, there has long been interest just in listening to sounds from Mars, in order to engage the public in planetary science. Wind blowing on another planet, rover sounds recordings, and even the Mars 2020 helicopter recordings are sounds of great interest for public outreach.

\begin{table}[h!]
\label{tab:various-noise}

\begin{tabular}{ | m{2.5cm} | m{2.5cm}| m{2.5cm} | m{2.5cm} |} 
  \hline
   \textbf{Potential Sound Source} & \textbf{Interest} & \textbf{Likelihood of success} & \textbf{Planning before launch} \\ 
  \hline
  \textbf{MOXIE}  & MOXIE Pumps behaviour  & Good & Coordination with MOXIE Team\\ 
  \hline
  \textbf{Mastcam-Z}  & Motors Monitoring during motion & Good & No\\ 
  \hline
  \textbf{Helicopter}  & Helicopter sound and video  & Weak (significant sound attenuation) & No \\ 
  \hline
  \textbf{Drill}  & Drilling process surveillance & Good & No\\ 
  \hline
  \textbf{Rover Wheels Noise}  & Interaction with soil  & Weak (microphone location not adapted) & No\\ 
  \hline
\end{tabular}

\caption{Possible sources of noise, with initial likelihood of success }
\label{tab:noise_source}
\end{table}

\subsection{Requirements summary}

\subsubsection{Science related requirements }

Due to its strong coupling with SuperCam LIBS investigation on rock hardness, we have designed the  microphone to be able to record the pressure wave generated by the LIBS shot, which has typical maximum amplitude of 5 Pa, at a distance of 4 meters from the rover mast, in a Martian atmosphere. In order to support the SuperCam geological investigation with classical signal processing methods, the signal-to-noise ratio (SNR) of the recording must be greater than 10 dB.

\begin{figure}[ht]
\noindent\includegraphics[width=0.9\textwidth]{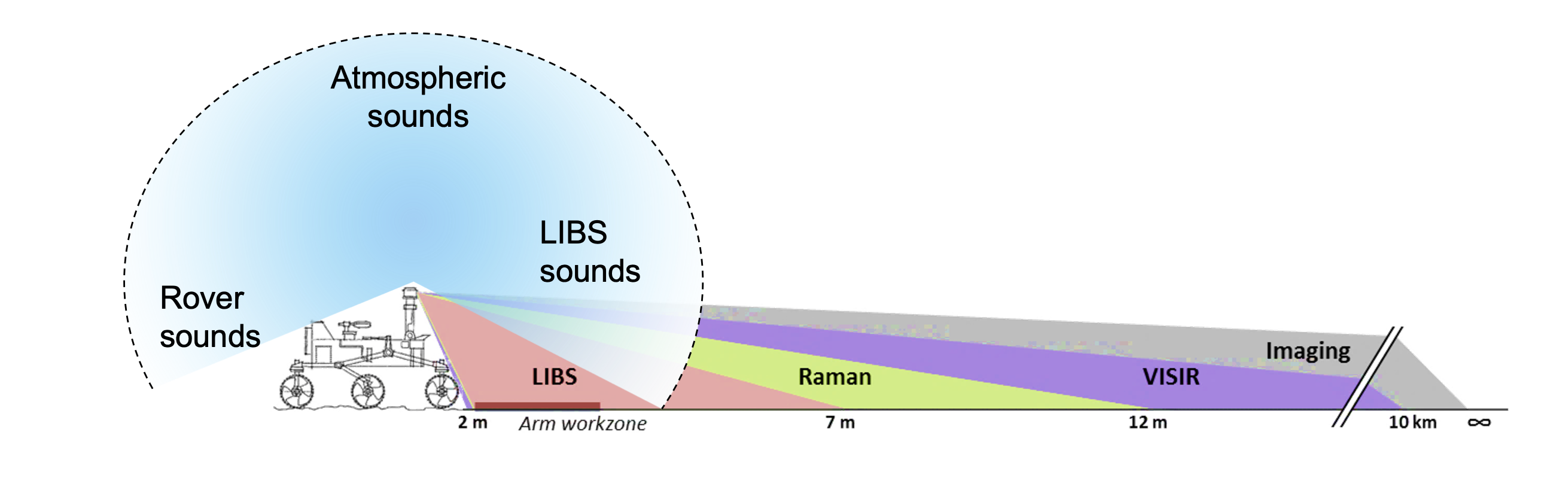}
\caption{Microphone range among the various SuperCam techniques. All techniques can operate at close range.}

\label{fig:SuperCam-Range}
\end{figure}

The expected bandwidth, as described in section \ref{LIBS-science} ranges from 100 Hz to 10 kHz. The optimal sampling frequency is 100 kHz to satisfy anti-aliasing criteria. A degraded mode allows a 25 kHz sampling frequency in order to save telemetry and increase the recording duration. The amplification gain must be tunable to be able to cope with various signal-to-noise ratios. This variability in the signal-to-noise ratio comes from the potentially small amplitude of the acoustic signals, possibly linked the variation in distance of the various targets (the requirement is to measure up to 4 meters), and to the variability in the LIBS acoustic counterpart depending on the target material properties, and on the background noise (wind). We have chose to keep the analog/digital (A/D) conversion dynamic higher than 60 dB to ensure the quantization effects are negligible with respect to the other error contributors.

As the Martian thermal environment is harsh, the microphone also includes a temperature sensor with relative accuracy of 1 K in order to compensate potential deviations in transfer function and noise. Test results (see \ref{Test_Campaigns}) have, however, demonstrated that the sensitivity to temperature is negligible with respect to the calibration capabilities, for both the microphone and its proximity electronics. 

The microphone design must also comply with the Martian wind as a potential source of perturbation with respect to the other signals of interest. The acquired signal must not be saturated by the effects of wind up to 1 $\sigma$ of the speed distribution as stated in the Mars 2020 Environment Requirements Document, which is 6 m/s. The test results (see section \ref{Test_Campaigns}) have shown that the saturation of the electronics occurs beyond a wind speed of 8 m/s and that the frequency content is limited to the lowest frequencies (inferior to 1 kHz), whereas the LIBS signal is mainly above 2 kHz. Usual filtering methods can then be used to separate those signals, provided that the saturation is avoided \citep{Murdoch2019}.

\subsubsection{Functional and design requirements}

The microphone was integrated late in the development of the SuperCam Mast-Unit. It was agreed with the Mars 2020 project that the microphone would have been descoped at any time if it had become a threat to other investigations. Fortunately, it survived the many challenges inherent to a space project. This specificity led to a strong design constraint: instead of having its own acquisition, the microphone "piggy backed" on existing the acquisition system of the SuperCam instrument, and uses the same A/D channel as the laser house-keeping. This led to minor operational constraints: the SuperCam team is not able to use the laser temperature housekeeping together with the microphone acquisition, and the total volume of an acquisition (and therefore its duration) is limited. 

The total amount of data for one acquisition cannot exceed 8 MB, which is the memory size allocated to one RMI (Remote Micro-Imager) image. This leads to a maximum time of recording of 41 or 167 seconds depending on the chosen sampling frequency. As part of the telemetry reduction effort, a filtering and decimation algorithm has been included in the SuperCam Body Unit (BU) flight software. The decimation factor is fixed to 4, while the 65-coefficient FIR filter ensures a rejection greater than 60 dB beyond 10 kHz.

The synchronization of the microphone recordings with the LIBS shots is mandatory, and is managed at higher level in the SuperCam Mast Unit. Thus, the amount of data necessary for the LIBS analysis can be reduced down to the minimum time window required to match the sound propagation delay and the LIBS signal duration constraints.  In order to save additional data, a specific "pulsed" mode has been introduced, in order to be able to record only the LIBS waveforms (as a consequence, this mode cannot be used to study wind turbulence).

The microphone is mounted beside the SuperCam telescope input window holder, in order to face in the direction of the the laser target, and detect the acoustic wave without any obstacle. The microphone is however omnidirectional for the lowest frequencies, even though it is limited by the surrounding large scale elements (Remote Warm Electronics Box (RWEB), Mast Unit, rover body, etc.) once integrated to the rest of the system.

The microphone, directly exposed to the Martian environment will undergo daily temperature variations from -80 °C to 0 °C and is qualified over a range from -135°C to +60 °C. The electronics, unable to operate properly at such low temperature, are attached to the Optics Box (OBOX) in order to be protected inside the Remote Warm Electronics Box surrounding the SuperCam instrument (RWEB), and is qualified over a temperature range from -55 °C to +60 °C. 

\subsubsection{Requirements flowdown }

\begin{figure}[ht]
\noindent\includegraphics[width=1\textwidth]{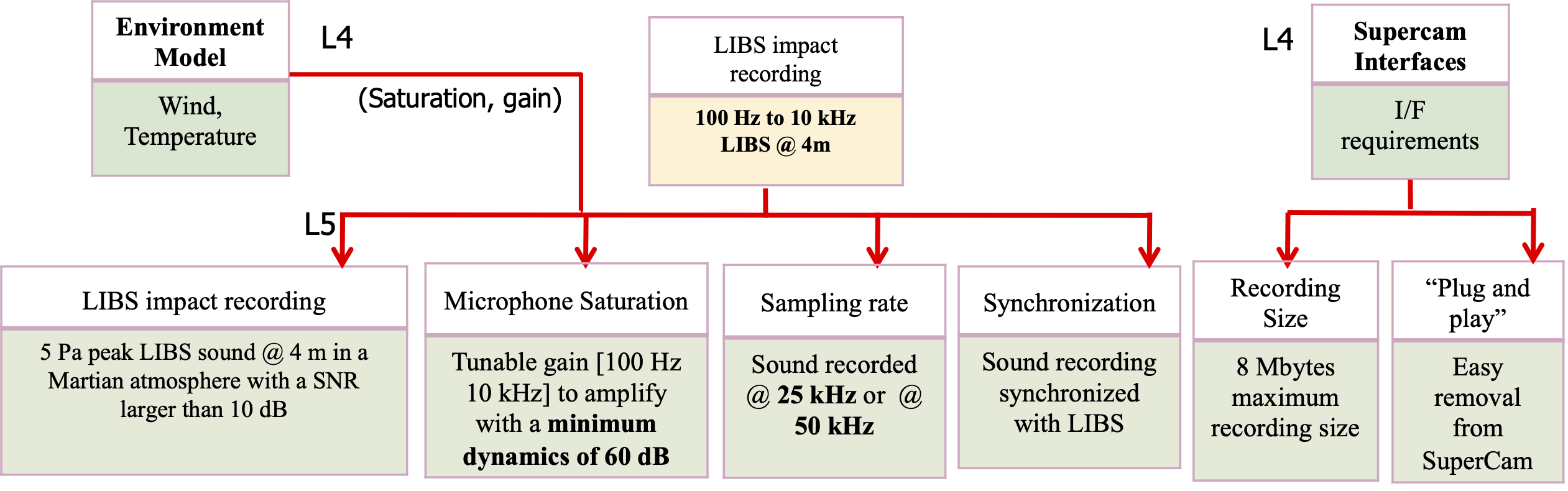}
\caption{SuperCam Microphone requirements flowdown. L4 and L5 are requirement levels 4 and 5}
\label{fig:requirements-flowdown}
\end{figure}

The summary of Science and design requirements has allowed us to setup the requirements flow-down for the microphone (Figure \ref{fig:requirements-flowdown}). The requirements for the SuperCam Microphone also included small mass and size, a rugged, robust design, and resistance to extreme conditions, including radiation exposure and low temperatures.

\section{Instrument design}

\subsection{Functional description}

\begin{figure}[h!]
\noindent\includegraphics[width=0.9\textwidth]{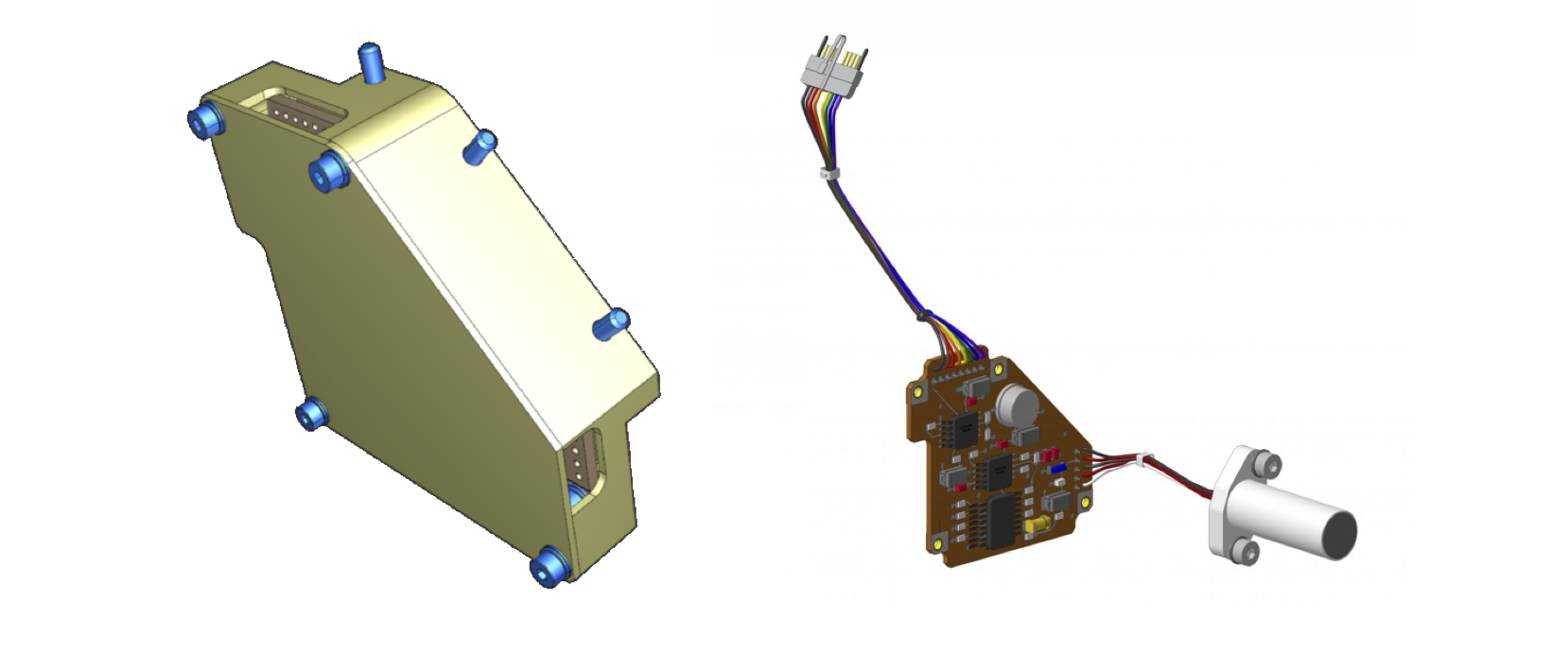}
\caption{Microphone subsystems overview. (Left) The Front-end electronics (FEE) box.  (Right) The FEE and the SuperCam Microphone cylinder or "finger".}
\label{fig:microphone_hardware}
\end{figure}

 The microphone is composed of two main parts: the microphone sensor located outside of the RWEB, and the front-end electronics (FEE) located inside the RWEB. The purpose of the FEE is to amplify the microphone output for acquisition by the housekeeping Analog-to-Digital Converter(ADC) of the instrument Digital Processing Unit (DPU). Figure \ref{fig:microphone_hardware} describes the various parts of the SuperCam Microphone.
 
The microphone sensor is exposed to the external environment, and is embedded in a cylinder required to pass through the RWEB. The microphone and the PT100 used for the temperature housekeeping are sealed in glue potting to avoid any unwanted motion.

\begin{figure}[h!]
\noindent
\includegraphics[width=0.8\textwidth]{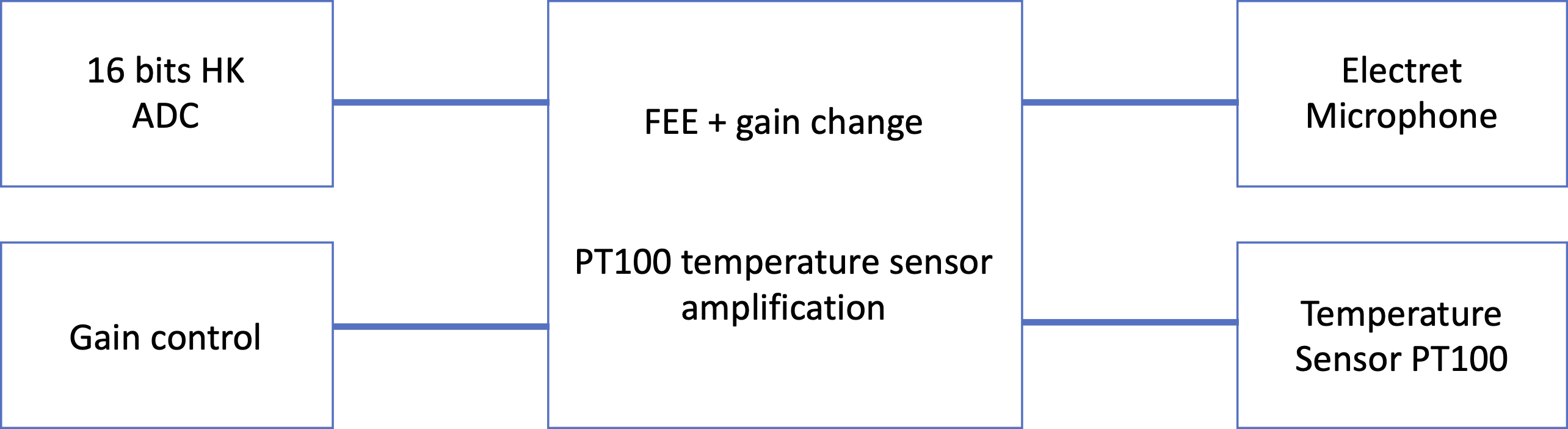}
\caption{Microphone functional overview. }
\label{fig:microphone_functional}
\end{figure}

Figure \ref{fig:microphone_functional} depicts the functional overview of the microphone subsystem. The sound wave (pressure) is converted into a voltage by the microphone finger outside the RWEB (Remote Warm Electronic Box). The resulting voltage is amplified by the Front End Electronics (FEE) located inside the RWEB. The FEE integrates a selectable gain (4 values) to match as closely as possible the input voltage range of the 12-bits laser housekeeping ADC, under any kind of signal amplitude. The recording is stored in the DPU memory, before being downloaded by the BU software, possibly compressed, and then stored in a non-volatile memory until the ground data download operations.

\begin{figure}[ht]
\noindent\includegraphics[width=0.5\textwidth]{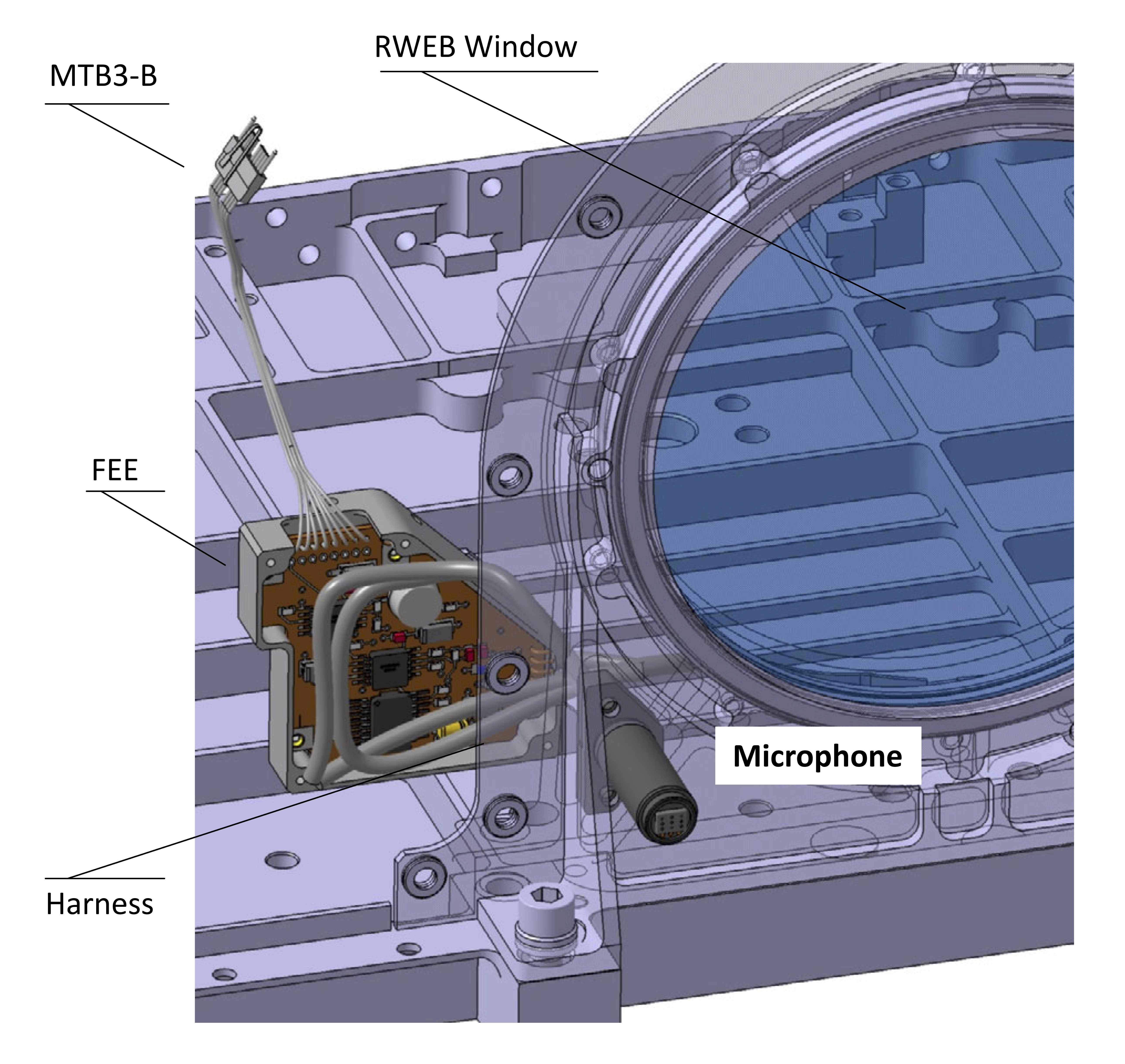}
\noindent\includegraphics[width=0.45\textwidth]{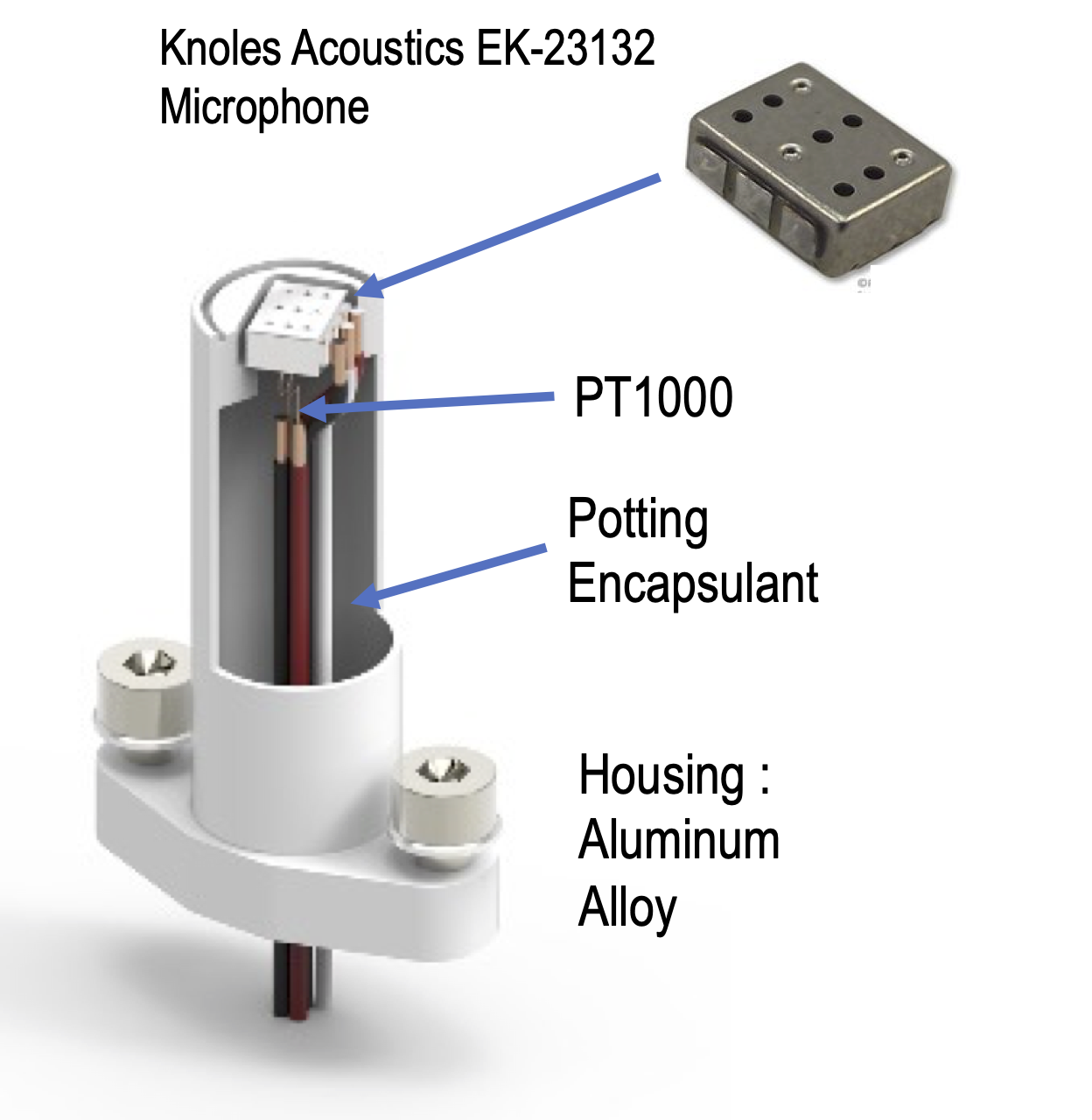}
\caption{(Left) Microphone finger implementation overview. (Right) Microphone assembly description. It encloses the microphone sensor, and a PT 1000 temperature sensor into an aluminum casing. The whole interior is filled with a cold-resistant E505 potting.}
\label{fig:microphone_accommodation}
\end{figure}

The electret microphone component (Figure \ref{fig:microphone_accommodation}) is a commercial off the shelf (COTS) component. It is the same commercial microphone sensor as used on the Mars Polar Lander and the Phoenix missions\citep{delory2007development, smith2004phoenix}. It sits outside the RWEB, at the tip of a 3 cm sandblasted aluminum "finger" (Figure 39). Hence the echo from the RWEB itself arrives between 222 µs and 277 µs after the direct signal. A temperature sensor is also potted inside the microphone stand. 

The temperature probe, a PT100 thermo-resistor, is powered by the same FEE, and the resulting voltage is amplified to provide a sufficiently large signal to the 16-bits ADC of the DPU, dedicated to the precise housekeeping data acquisition of the SuperCam Mast Unit. The microphone temperature data is stored together with the other SuperCam MU housekeeping data.

A shielded cable connects the microphone and temperature sensors to the FEE inside the OBOX (Figure \ref{fig:microphone_accommodation}). The harness consists of five Manganin wires to limit the thermal leak. When the electret microphone is at -120°C and the OBOX at -35°C, the power drawn from the survival heaters by the microphone is only ~30 mW, four times smaller than for standard copper wires of the same diameter.

The role of the FEE is to amplify and filter the analogic signal and to collect temperature data. A first stage amplifies the signal by a factor x15, a second stage by a controllable gain, x2, x4, x16, and x64. The digitalization of the signal is performed by a fast 12-bit ADC on the instrument DPU board. To avoid failure propagation, two short-circuit protections are implemented for the microphone and the operational amplifiers. 

We had initially considered to implement a grid to protect the microphone sensor from Martian dust, but the decrease in sensitivity of the overall assembly when the grid was present led us to favour the performance and remove the grid from the design.


\subsubsection{Microphone component properties}
\label{Microphone_component}

The SuperCam microphone is a Knowles EK-23132 microphone (Figure \ref{fig:microphone_accommodation}). This is an electret based sensor, using a charged membrane, whose movement due to pressure fluctuations alters the capacitance of the sensor, which is then read as a signal. The EK-23132, originally selected for the NASA Mars Polar Lander (MPL) \citep{delory2007development}, is the lowest noise microphone manufactured by Knowles, and in our experience has a sensitivity superior to similar microphones made by other manufacturers. It is designed to be inherently rugged to withstand severe environmental conditions, and has a low vibration and shock sensitivity. The microphone contains a Bipolar Junction Transistor (BJT) that amplifies the charge variations caused by the membrane motion, transforming this signal into a voltage level through an output bypass resistor.

EK-23132 sensitivity is 29.6 mV/Pa at 1 kHz without any stage of amplification. Its dimensions are 5.6 mm x 3 mm. 

\begin{center}
\begin{figure}
\includegraphics[width=0.8\textwidth]{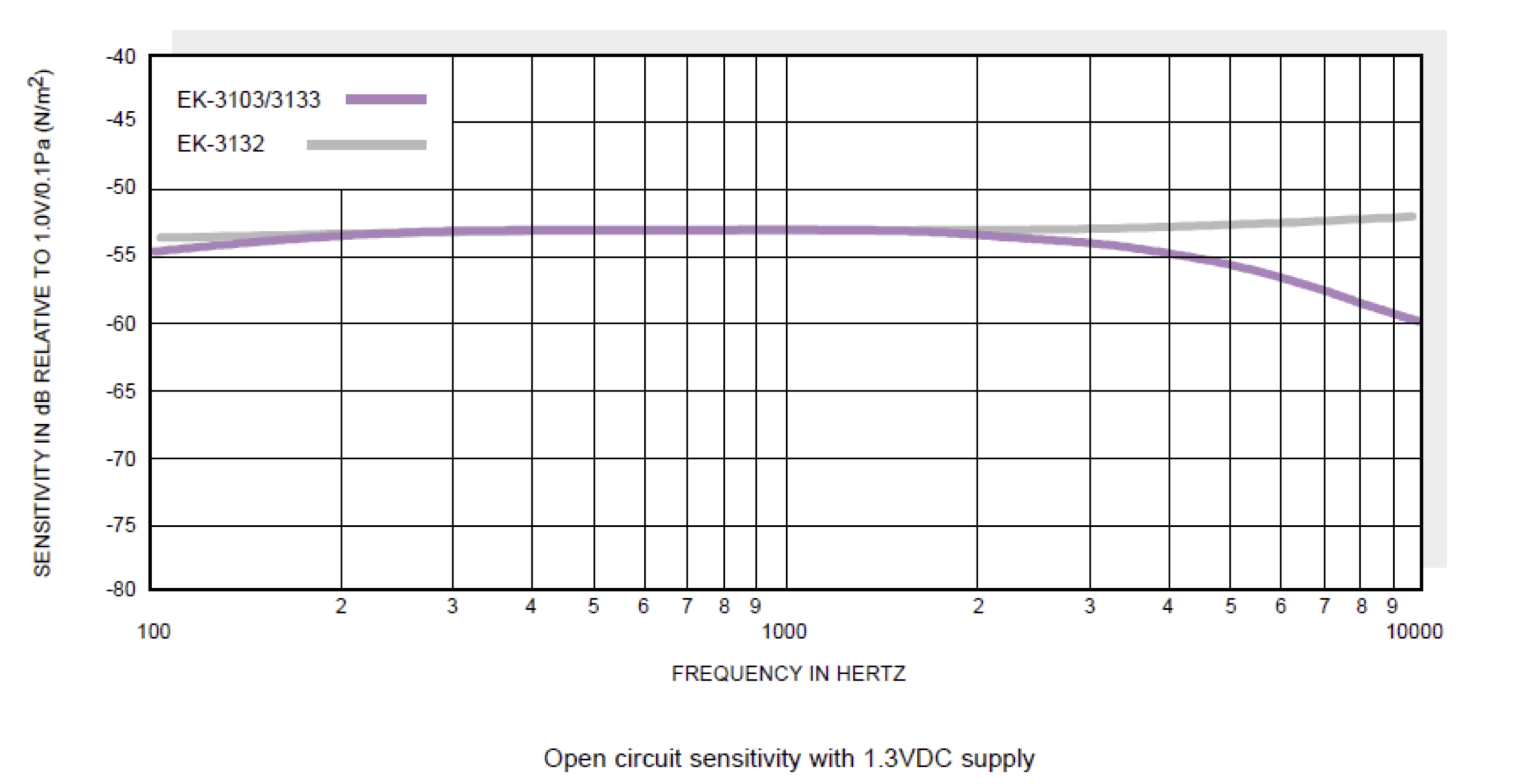}
\caption{Microphone sensitivity. The EK-3132 underwent a full qualification process, and was tested for its performance at low atmospheric pressures, low temperatures, and for radiation resistance. Low pressure tests over a variety of temperatures were conducted in air using standard thermal vacuum chambers. Figure from EK3132 datasheet.}
\label{fig:EK3132-sensitivity}
\end{figure}
\end{center}

Figure\ref{fig:EK3132-sensitivity} shows the EK-23132 sensitivity and frequency response. Theoretically, the sensitivity of the microphone scales as the acoustic impedance $\rho c$, where $\rho$ is the density of the gas and $c$ the sound speed. $\rho c$ for Mars is $\sim 0.01 \rho c$ for Earth, [Sparrow, 1999], thus reducing $\rho$ by 100 for constant $c$ in air achieves a similar effect.

\begin{figure}
\includegraphics[width=0.8\textwidth]{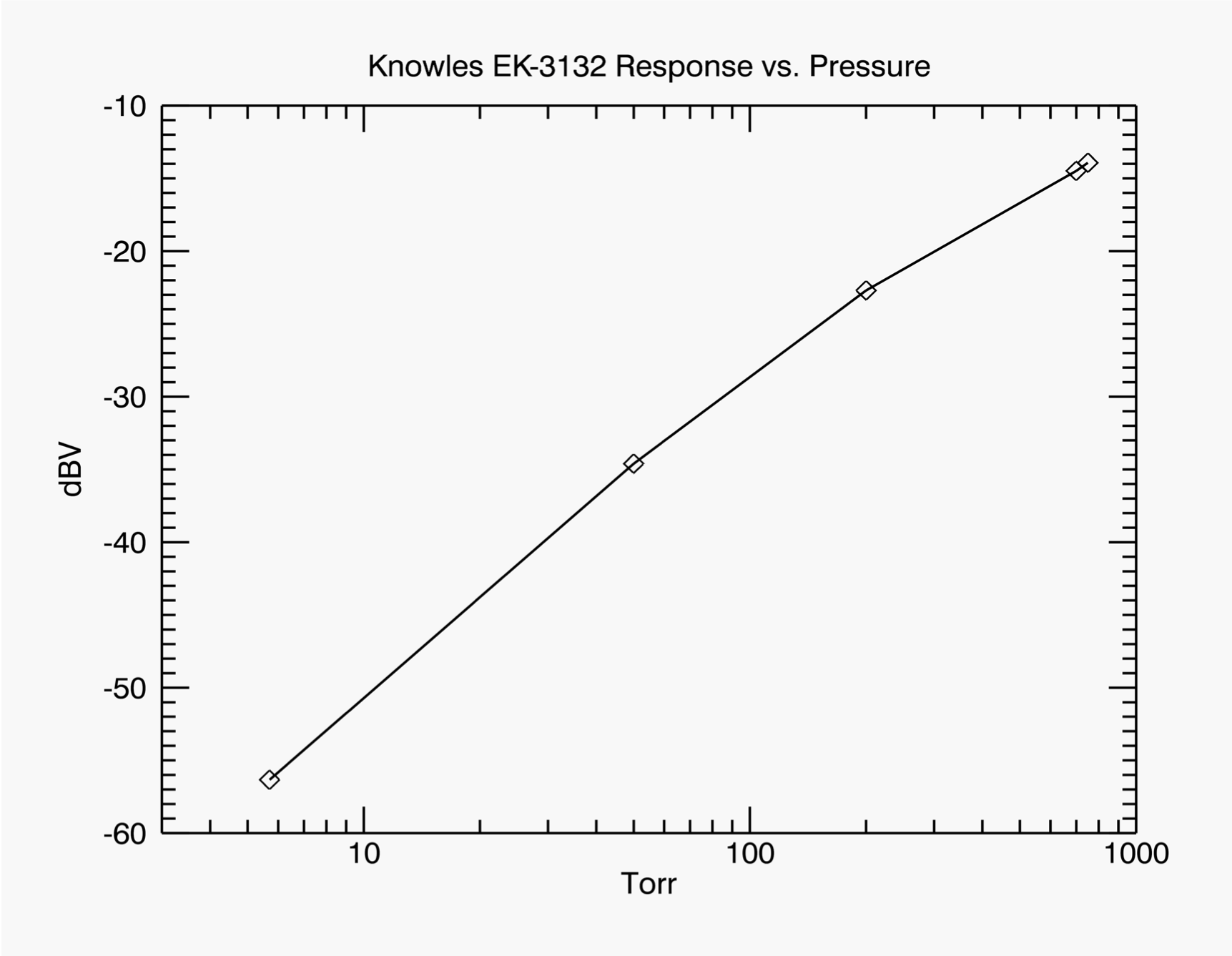}
\caption{Output in dBV (decibels relative to 1V output) for the EK-23132 as a function of atmospheric pressure in Torr - from EK3132 datasheet}
\label{fig:EK3132-pressure}
\end{figure}

\subsubsection{Proximity Electronics}

The microphone front-end electronics (FEE) has four functions
\begin{enumerate}
\item Amplification of the microphone signal
\item Gain switch (2, 8, 32 and 64)
\item PT100 Temperature resistance to voltage conversion 
\item Microphone power supply +3.3 V
\item Interface with the O-BOX
\end{enumerate}

The  FEE ensures the polarization of the components and the amplification of the microphone signal. Amplification is done with two stages. The first amplification is fixed, the second amplifier has a selectable gain of 2, 4, 16 or 64 and a bandwidth from 100 Hz to 10 kHz. High gains (32 and 64) have been designed to optimize the SNR with respect to the LIBS sound recordings made in lab. Low gains (2 and 4) are meant to record environmental noise while avoiding saturation due to e.g., wind gusts.

\begin{center} 

\begin{table}[h!]
\begin{tabularx}{0.8\textwidth} { | >{\raggedright\arraybackslash}X | >{\centering\arraybackslash}X | >{\centering\arraybackslash}X| >{\centering\arraybackslash}X | }

\hline
\textbf{Gain Number} & \textbf{Measured FEE Gain [V/V]} & \textbf{Total Sensitivity [V/Pa]} & \textbf{Total Sensitivity [LSB/Pa]} \\
\hline
\textbf{Gain 1} &  29 & 0.6 &  491\\
\hline
\textbf{Gain 2} &  57 & 1.2 &  983\\
\hline
\textbf{Gain 3} & 240 & 5.2 &  4262\\
\hline
\textbf{Gain 4} & 972 & 21.0&  17213\\
\hline
\end{tabularx}
\caption{Gains of the microphone electronics and conversion to physical units . This table includes all amplification gains - fixed and tunable.}
\label{tab:MIC-GAINS}
\end{table}
\end{center} 

The FEE is directly controlled by the SuperCam power unit (DPU). The measured total gain and the total sensitivity of the microphone flight model at 1 kHz and its electronics are presented in the Table \ref{tab:MIC-GAINS}.  These are the begining-of-life values but, as mentioned above, we do expect some evolution in the microphone sensitivity over time. The FEE also provides the output of the PT100 temperature probe. The total sensitivity of the temperature measurement is 3.96 mV/K. The output voltage is shifted by a nominal offset of 1.029 V at 0 °C. 

The whole design has been made keeping in mind the protection of the SuperCam main electronics. Therefore a protection against failure, shorts, saturation or single event transient (SET) has been implemented at the input of the FEE. The total power consumption of the microphone FEE is about 20 mW, on $+/- 5V$. Figure \ref{fig:microphone_hardware} shows the flight model of the FEE during its final inspection before delivery. The FEE box mechanical design is very straightforward: a simple 0.5 mm aluminium box enclosing the FEE (Figure \ref{fig:microphone_hardware}, left). The FEE box is connected to the chassis mass.  

\begin{center}
\begin{figure}
\includegraphics[width=0.6\textwidth]{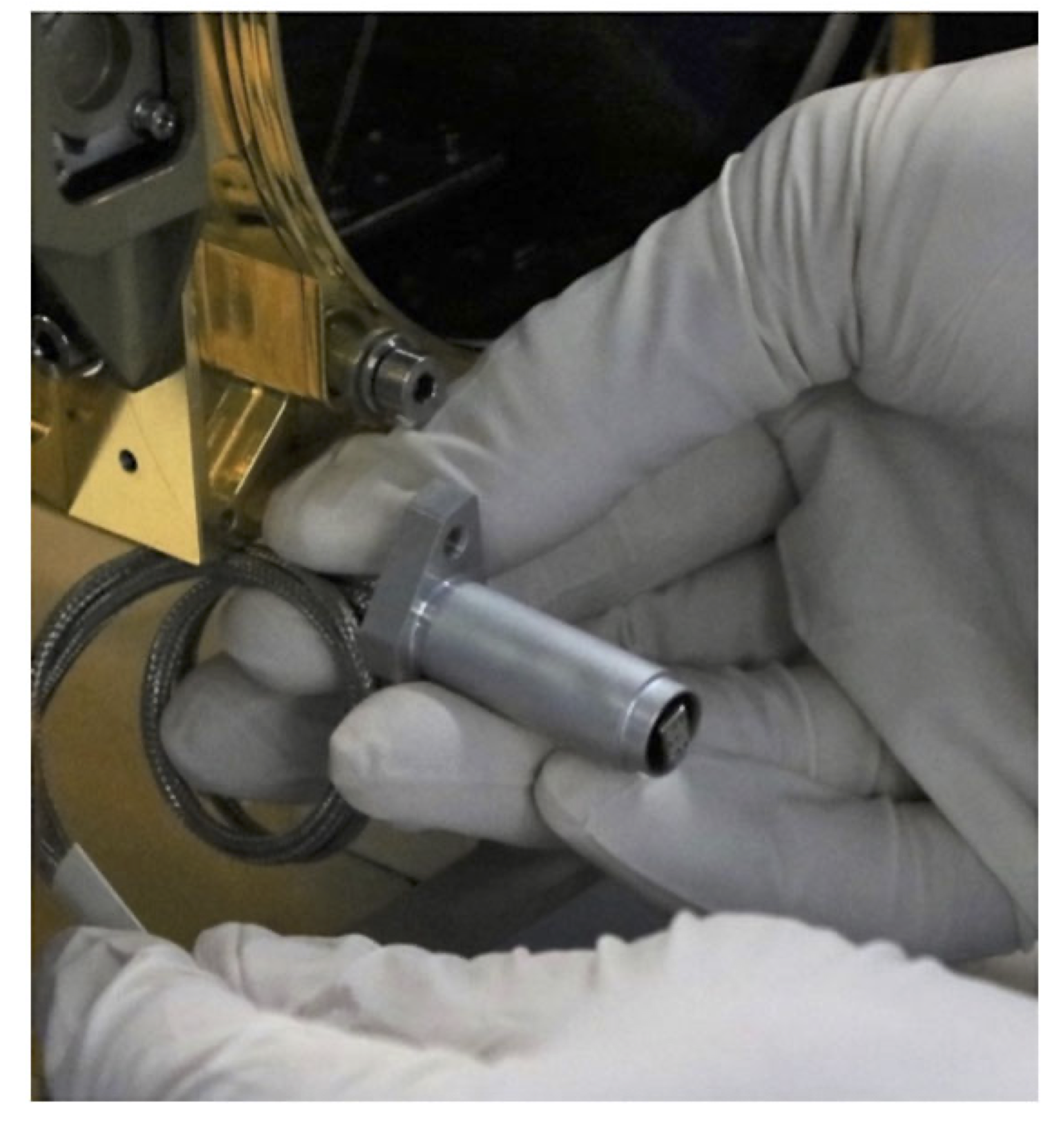}
\caption{Microphone hardware during integration}
\label{fig:Microphone_hardware}
\end{figure}
\end{center}

\subsubsection{Microphone Assembly Thermal design}

The temperature range of the FEE is standard: the parts and process selected allow the design to cope with the required temperature range. However, the MIC temperature range is extended down to -130°C (lowest possible temperature on Mars), therefore a dedicated qualification has be done for the microphone assembly. Due to the external location of the microphone, the thermal architecture has to minimize the microphone thermal leaks. If a standard strategy using cables and shielding was applied, the conductance of the cable between the microphone assembly and its FEE  would be to high to cope with instrument safe mode heating power requirements.  We have, therefore, chosen to implement a thermal leak reduction for the cable, thanks to "athermous"  Manganin \textsuperscript{TM}  wires, an alloy that conducts current but has a low thermal conductivity.

\subsubsection{Data handling}

As required by the instrument design, the SuperCam Microphone is mostly managed by the SuperCam Mast-Unit. The SuperCam Body-Unit is mostly in charge of implementing the data storage, as well as the interface with the Rover. The Mast-Unit controls various subsystems, such as the laser,  autofocus, Microphone, IRS, RMI, and SDRAM. 

\begin{figure}
\includegraphics[width=0.9\textwidth]{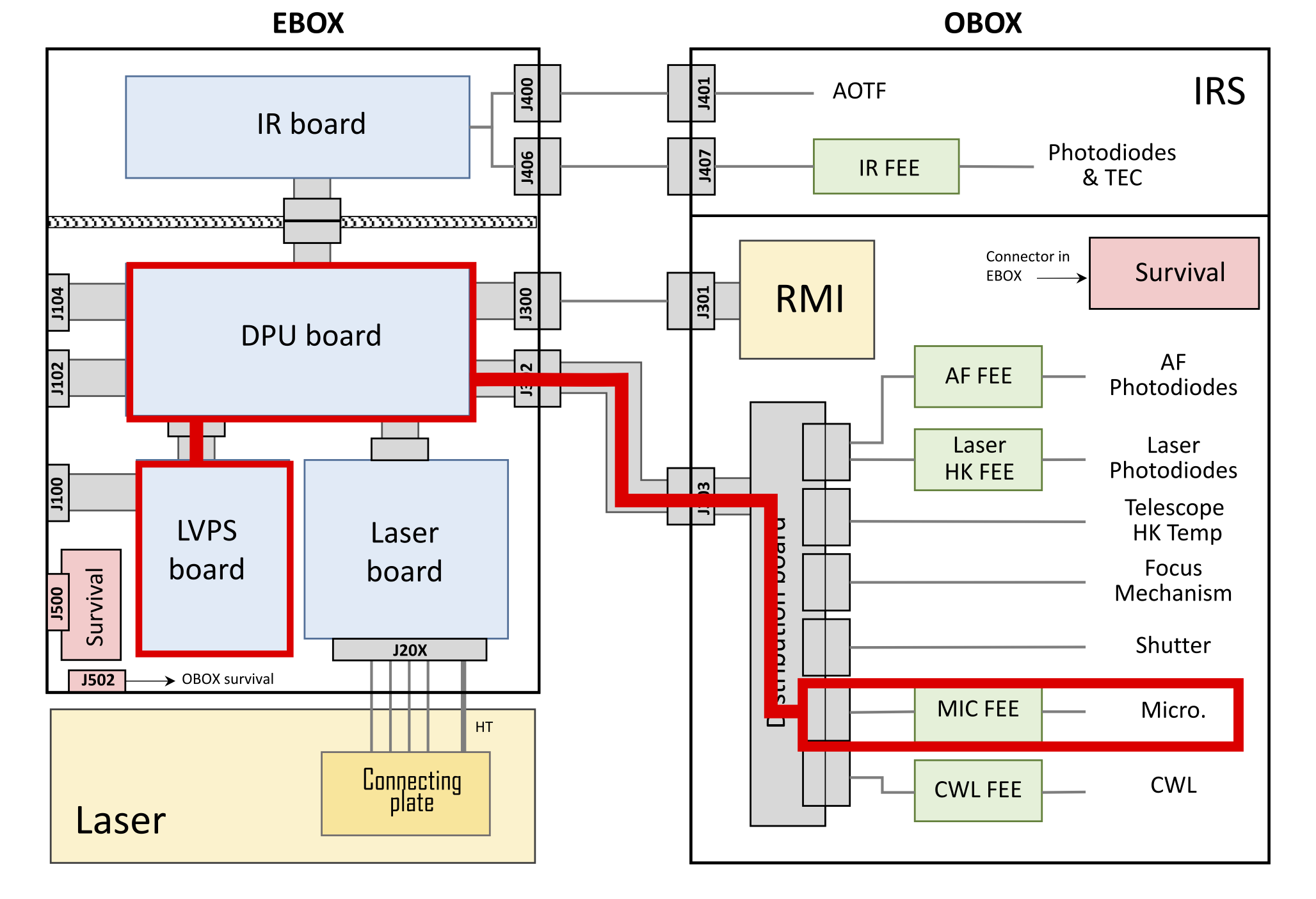}
\caption{Microphone Data Handling Architecture. The Microphone data processing takes mainly place in the SuperCam Mast-Unit. Figure modified after\citep{Maurice2021}.}
\label{fig:MIC-DATA-HANDLING}
\end{figure}

The  Microphone driver derives from the laser HouseKeeping driver which drives the high-speed ADC (sampling frequency up to 100 kHz) and stores data. This driver is also called by the laser driver to synchronize the recording to the LIBS shots when needed. The DPU Unit controls the microphone. Due to the SuperCam Body Unit software limitations, the recording limited to a 8 MB data product size.

\begin{figure}[ht]
\includegraphics[width=0.9\textwidth]{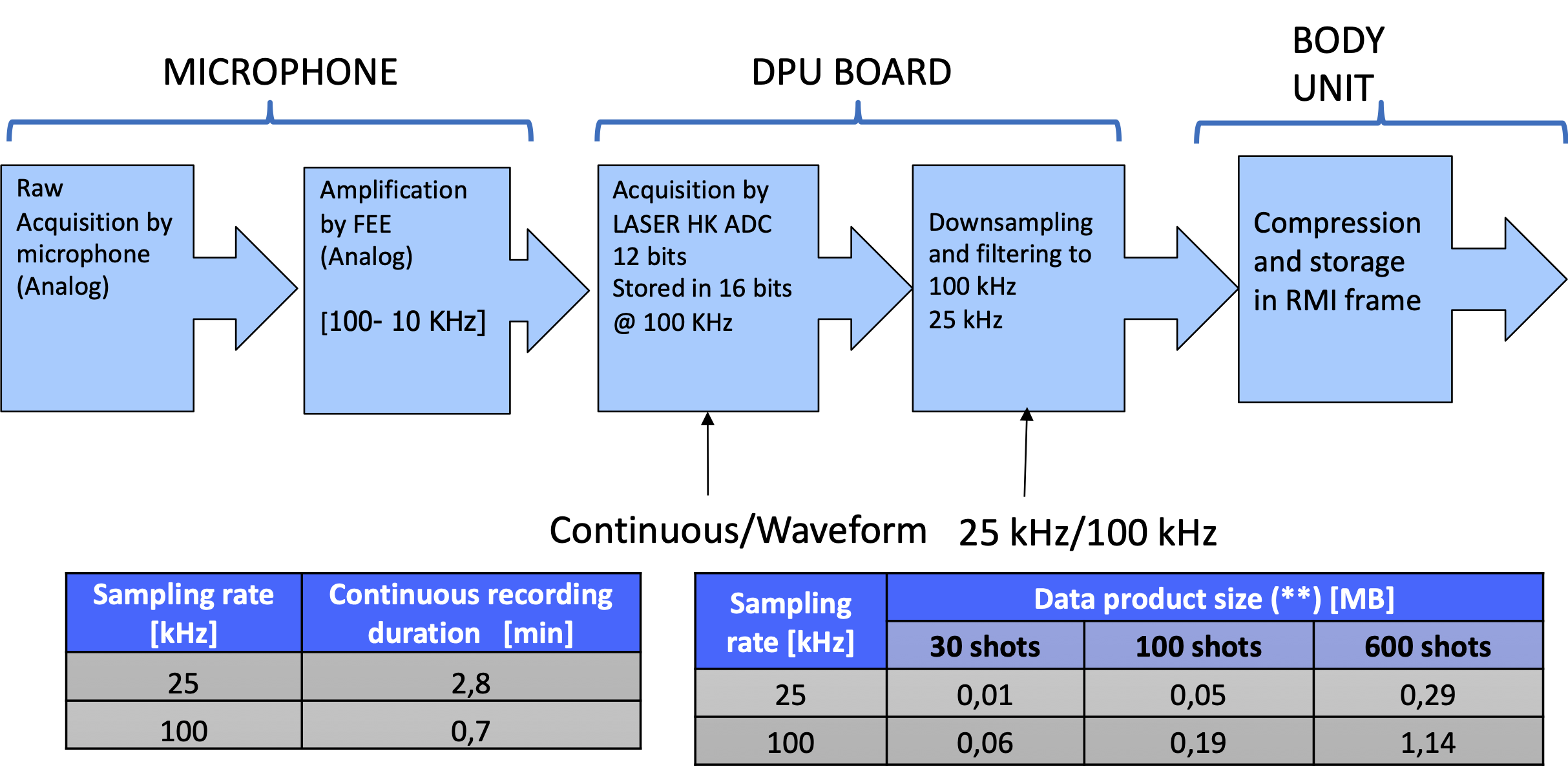}
\caption{Microphone data processing. A schematic overview of the data processing. The sampling rate and duration depend on the operational mode of the microphone (see Section \ref{sec:operation}).}
\label{fig:MIC-DATA-PROCESSING}
\end{figure}

\subsubsection{Operation Modes}
\label{sec:operation}

\begin{figure}[ht]
\includegraphics[width=1\textwidth]{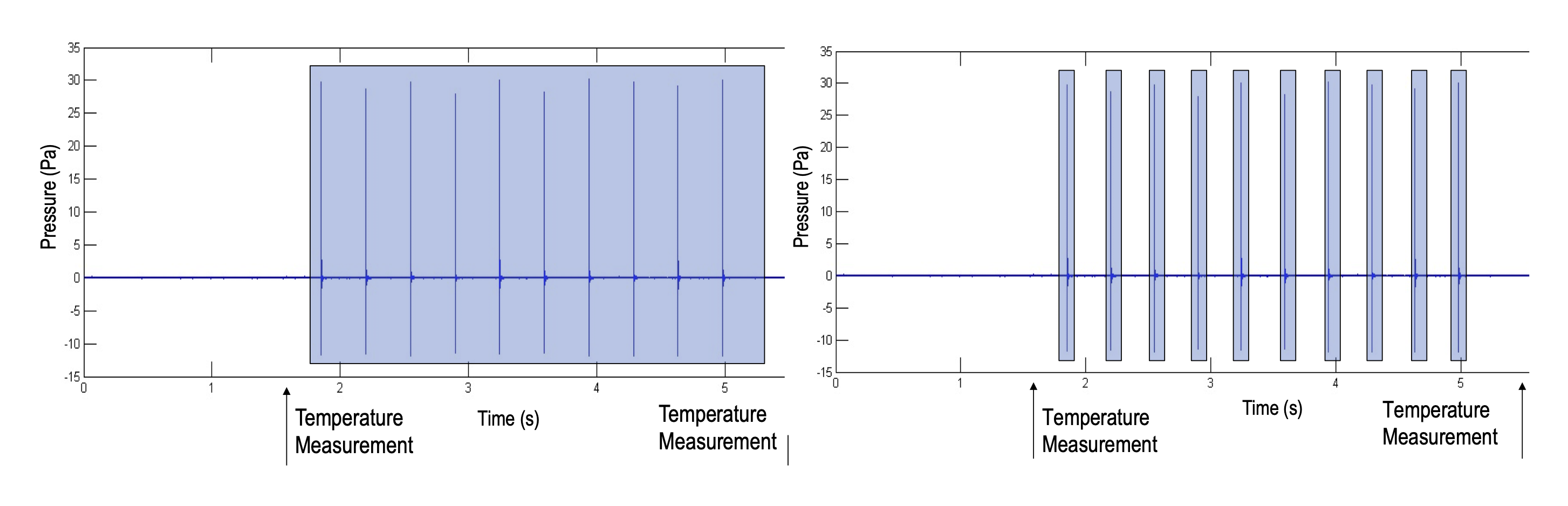}
\caption{Graphical representation of MIC standalone / MIC + LIBS continuous mode / MIC + LIBS pulsed mode}
\label{fig:MIC-rec-time}
\end{figure}

\begin{figure}[ht!]
\noindent\includegraphics[width=0.9\textwidth]{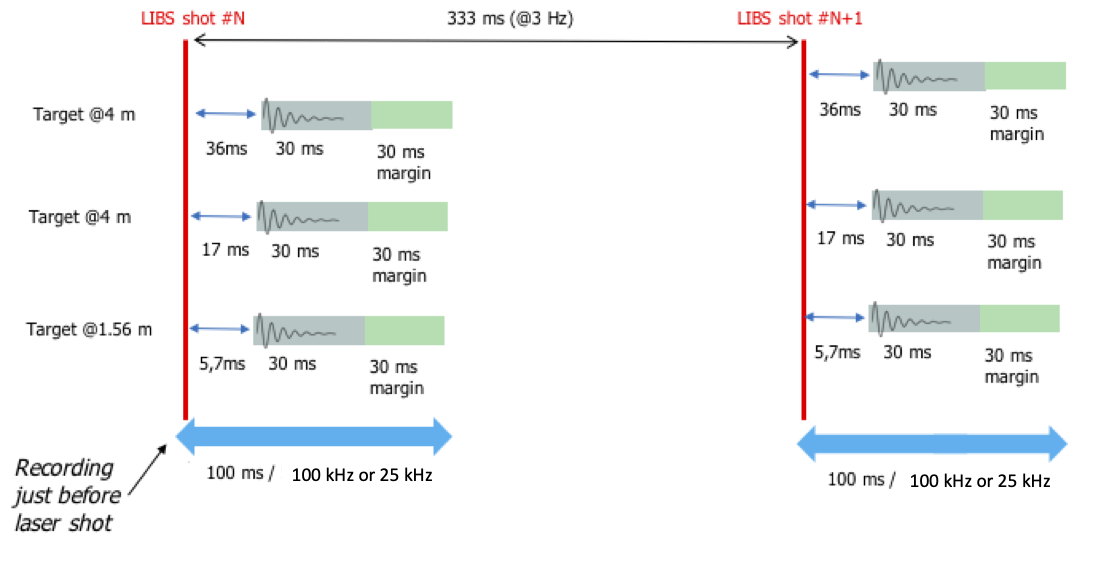}
\caption{Microphone pulsed mode timing. The cycle is based on the SuperCam laser raster of 333 ms. The pulsed mode will record 100 ms every 333 ms. This will allow the timing to accommodate all possible distances of the SuperCam LIBS target. }
\label{fig:PulsedModeTiming}
\end{figure}

The Microphone has three  modes of operation (see also Figure \ref{fig:MIC-rec-time}):
\begin{enumerate}
\item	 \textbf{MIC standalone}: this mode is mostly dedicated to the study of natural (winds) or artificial  (Rover, helicopter ...) sounds. It can also be used in coordination with other payloads such as the MastCam-Z or MOXIE.  
\item	\textbf{MIC + LIBS continuous mode}: used to record sound continuously during a LIBS burst.
\item	\textbf{MIC + LIBS Pulsed mode}: used to sample specifically the LIBS burst. The sound recording starts less than 1 ms before the laser pulsed is emitted and runs during 60 ms for each shot (except the last one kept for laser data). It allows to keep data only related to the study of the LIBS shots. The timing for the pulsed mode is shown in Figure \ref{fig:PulsedModeTiming}.
\end{enumerate}

For these 3 modes, we can use the three possible gains (from 0 to 3). Due to storage capability, the recording duration is, at most, 167s for a sampling frequency of 25 kHz, and 41s for a sampling frequency of 100 kHz. An optional decimation algorithm is also implemented in the Body-Unit to down sample the data from 100 kHz to 25 kHz. We anticipate that the gain settings will depend mostly on environment and target distance.

\section{Performance Model}

\subsection{Microphone model}

 The frequency response of the microphone is established with respect to the manufacturer datasheet. However, to perform an extended analysis of the instrument, it is necessary to consider a larger frequency band (at least from 10 Hz up to 100 kHz). The microphone was then modeled with a first order high-pass filter $(f_{HP, M}=30 \mathrm{~Hz})$  and a second order low-pass filter $(f_{LP, M} = 15 kHz, h_M = 0.2)$, and adjusted in amplitude (Microphone sensitivity = 22.4 mV/Pa). The resulting transfer function is described by Equation \ref{transfer-function}
\begin{equation}
    H_{n}(f)=S_{M} \frac{j\left(\frac{f}{f_{HP, M}}\right)}{1+j\left(\frac{f}{f_{HP, M}}\right)} \frac{1}{1+2jH_m\left(\frac{f}{f_{LP, M}}\right)-\left(\frac{f}{f_{LP, M}}\right)^2} 
\label{transfer-function}
\end{equation}

\noindent and compared to datasheet values. The model is justified at low and high frequencies by the typical sensitivity of similar products of Knowles Electronics. However, without the real damping factor and the cutoff frequency, only likely assumptions have been made so far. The resonance above 10 kHz might have a more complex description than a second order response, due to the microphone inlet. This is particularly important for the phase modelling and filtering considerations.

\begin{figure}[ht!]
\includegraphics[width=1\textwidth]{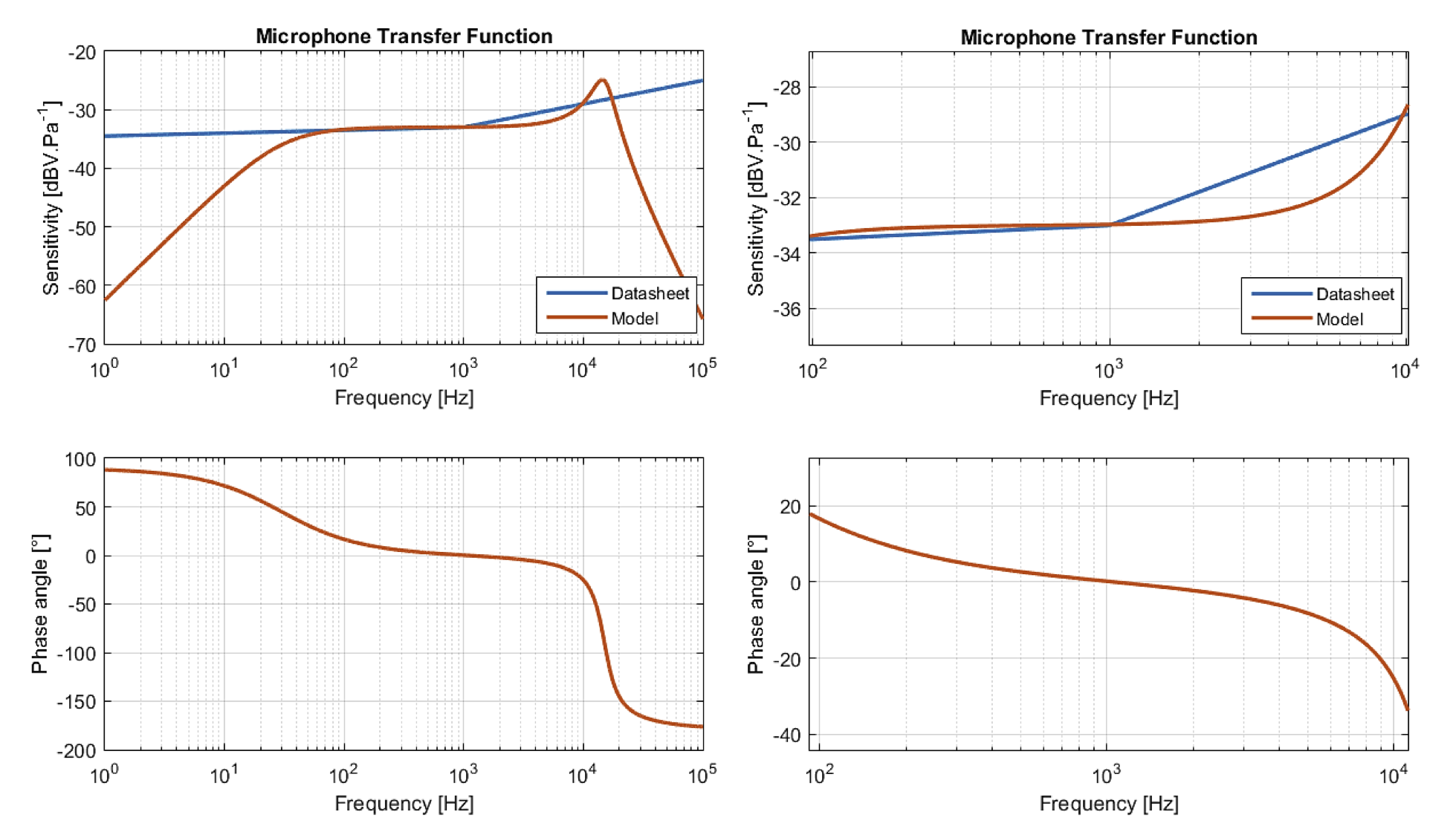}
\caption{Theoretical Transfer functions of the microphone model compared to the datasheet. (Left) Full bandwidth. (Right) Zoom on the band 100 Hz – 10 kHz.}
\label{fig:EK-3132-transfer-function-model}
\end{figure}

\subsection{FEE Transfer function models}

\begin{figure}[ht!]
\includegraphics[width=0.8\textwidth]{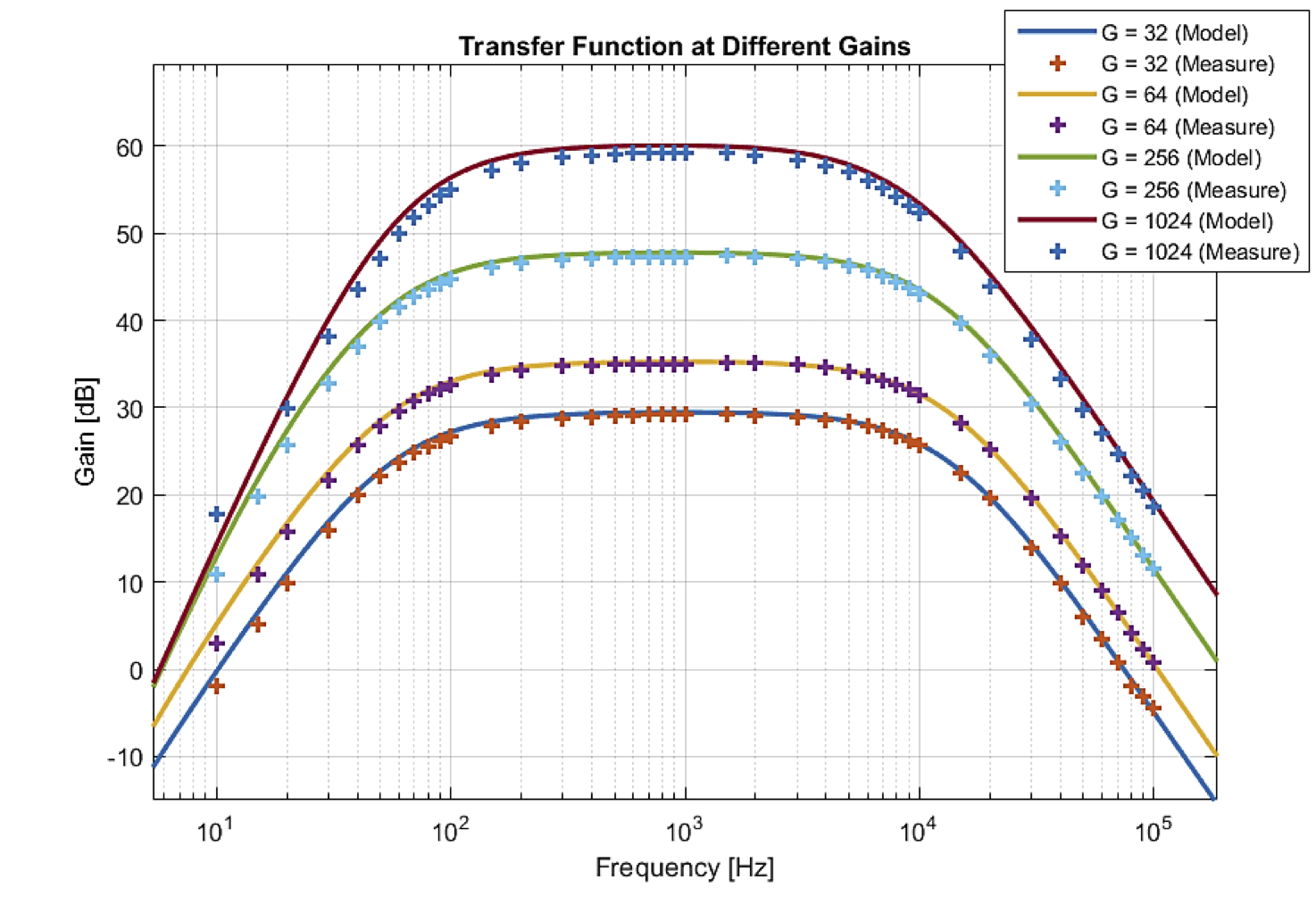}
\caption{Comparison of the models (solid lines) with the measurements (+ symbols) of the amplification electronics transfer function (amplitude only).}
\label{fig:FEEtransfer-function-models}
\end{figure}

The electronic transfer function is the response of two 1st order band-pass filters, plus the output stage representing the behavior of the level shifter (+2.5 V reference and associated passive components) and a high impedance input (buffer or oscilloscope). It is modelled by including the effect of the non-perfect operational amplifiers. The comparisons of the theoretical results with the measurements are presented figure \ref{fig:FEEtransfer-function-models}. The model is able to reproduce the transfer function amplitude with a precision inferior to 1 dB.

\subsection{Overall System Noise}

\begin{figure}[ht]
\includegraphics[width=0.7\textwidth]{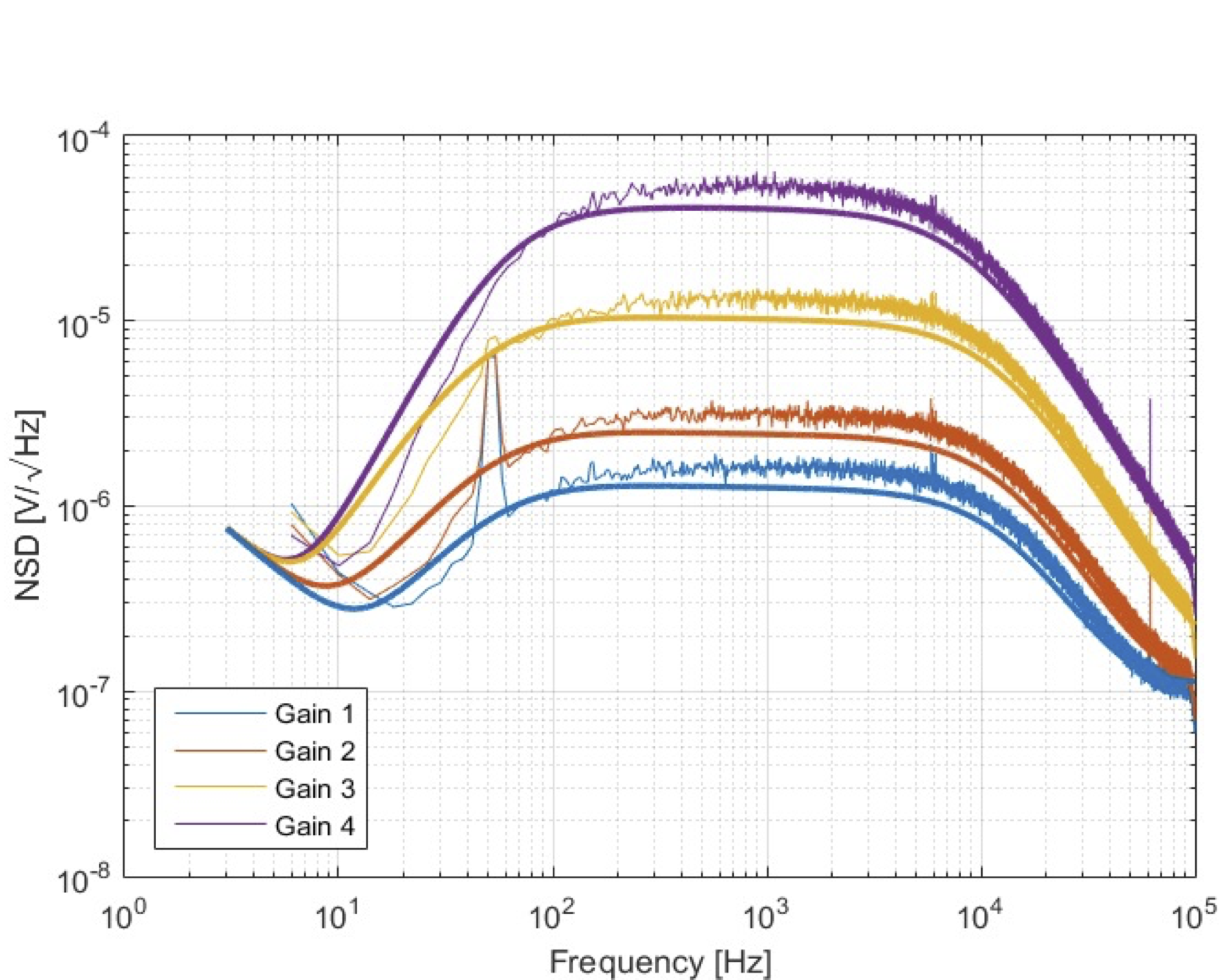}
\caption{Comparison between the front-end electronics noise (amplitude) spectral density models (plain lines) and the measurements (noisy curves) for all gains at +25 °C. Note that the remaining 50 Hz peak is due to the difficulty obtaining a clean EMC environment in the lab.}
\label{fig:FEE-Noise}
\end{figure}

The electronic noise is measured with an oscilloscope as the root mean square value of the signal, when no signal is applied at the input of the circuit. Theoretical values computed with the model and measurements are presented in Figure \ref{fig:FEE-Noise}. The order of magnitude of the noise is well reproduced  by the model. The remaining small discrepancies are due to the simplifications made in our model. For comparison, an ADC LSB with an input voltage range of 5 V and 12 bits of resolution is about 1.2 mV. Therefore, the lowest gains (Gains 1 and 2, see Table \ref{tab:MIC-GAINS}) can be used to improve the maximum input range without saturating the ADC, whereas the highest gains (Gains 3 and 4, see Table \ref{tab:MIC-GAINS}) are more relevant for the recordings of low amplitude signals and the characterization of background noises.

\subsection{Theoretical SNR for LIBS}

\begin{figure}[ht!]
\includegraphics[width=0.65\textwidth]{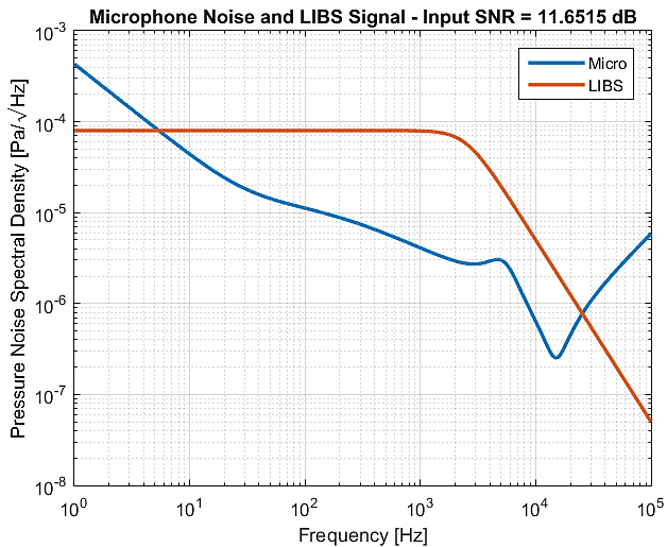}
\caption{Comparison between the equivalent input noise and the LIBS signal.}
\label{fig:MIC-SNR}
\end{figure}

The complete transfer function amplitudes (sensitivities), including the microphone plus the front-end electronics, are presented in Figure \ref{fig:MIC-SNR}. Each curve corresponds to a defined gain of the second stage of amplification. The LIBS signal amplitude shape used has been derived from recordings made in the lab with calibrated microphones.

\section{Microphone Verification Process  }
\label{Test_Campaigns}

\subsection{Microphone part verification}

Given that the microphone is a commercial off-the-shelf (COTS) component, its integration into a space instrument development required some additional qualifications. A batch of a hundred components was procured from the same manufacturer sub-lot, in order to perform all the necessary tests and integration processes.

First, an entrance visual inspection was performed to verify the lot and sub-lot numbers of each part. Second, a serial number marking operation was performed to provide a non-ambiguous traceability of each part. Then an initial performance test was performed to track potential degradation during further operations by comparison. The test was based on noise, bias, and sensitivity measurements in a contained acoustic bench, which is dedicated to the whole process. The main group of components (71 parts) underwent vibration, pin test and re-tinning operation, whereas two others groups were dedicated respectively to radiation and destructive physical analysis (DPA). After the main group performances verification, five parts were finally extracted for internal X-ray inspection and surface electron spectroscopy of the solder pads. No physical or performance degradation was observed at this step. The 20 best components in terms of acoustic sensitivity were then selected to be integrated in the microphone assembly. Six were allocated to PQV (package qualification and verification) testing and further inspections, the five best assemblies were used for the final qualification thermal test of the models eligible for flight.
\begin{figure}
\begin{center}
\includegraphics[width=0.6\textwidth]{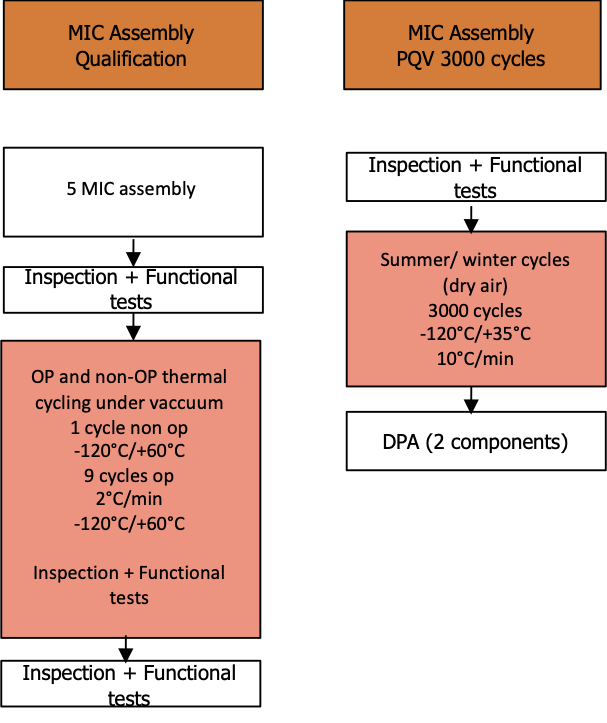}
\caption{Microphone test flow summary. Two tracks were used:  (left) a track for the qualification, and (right) a track for the complete package qualification and verification (PQV).}
\label{fig:MicrophoneTestFlow}
\end{center}
\end{figure}
\begin{figure}[ht!]
\begin{center}

\includegraphics[width=.5\textwidth]{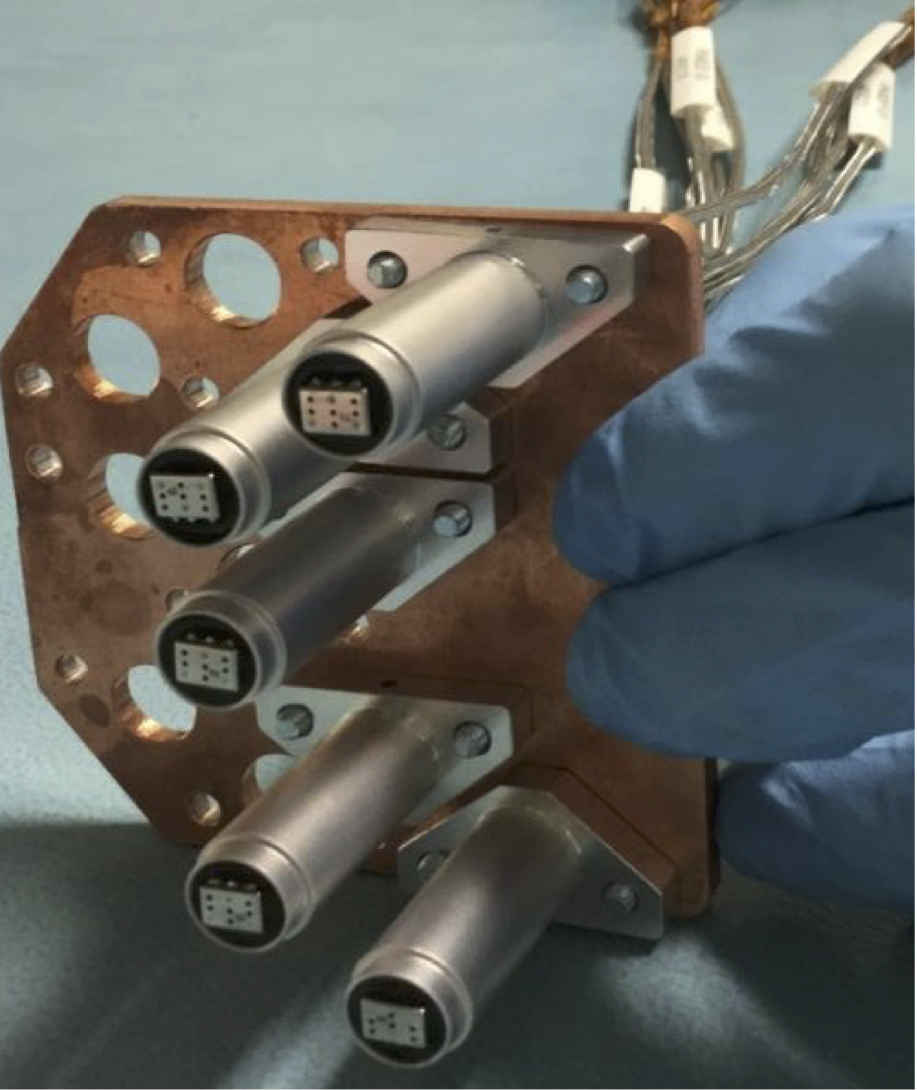}
\caption{Picture of the 5 microphone assemblies mounted on the copper interface for cryogenic temperature qualification.}
\label{fig:MIC-CRYO}
\end{center}
\end{figure}
Contrary to the rest of the SuperCam Mast Unit instrument, the microphone is directly exposed to the Martian environment. As a consequence, six microphone assemblies have undergone a package qualification and verification process (PQV) that consist of 1400 thermal summer cycles (+40 to -105 °C) and 600 winter cycles (+15 to -130 °C). No major potting lift-off was observed that would endanger the cleanliness or the sealing of the RWEB, or the microphone integrity itself. No degradation of the acoustic performance was observed during the whole campaign. The PQV was performed at CNES in a thermal chamber supplied with liquid nitrogen.

A sub-lot of five microphone assembly parts eligible for flight was tested for a cryogenic temperature qualification at ISAE-SUPAERO in a dedicated thermal vacuum chamber operating with a liquid nitrogen thermal exchanger. The parts were mounted on a copper mechanical interface, as shown in Figure \ref{fig:MIC-CRYO}, and the temperature of each assembly was monitored thanks to its internal PT100 sensor. Four cycles were performed from +60°C to -135°C: two of them under cruise pressure conditions (10$^{-5}$ mbar), the microphone being off, and two under Martian conditions (5 to 10 mbar) with a functional test on the dwells. The performances of each assembly have been compared to their pre-qualification characteristics with the acoustic test bench. No measurable differences were observed and the five microphone assemblies were therefore declared qualified for the mission thermal environment. The flight model and its spare were picked from this sub-lot.

\subsection{Radiation Susceptibility}
Screening and lot qualification were performed on 100 microphone parts. The qualification path included detailed component analyses, and radiation sensitivity evaluation, with increasing ionizing doses up to 1750 rad (Si). This maximum dose results is a loss of 10 dB of sensitivity for the full mission (Radiation Design Factor of 2). This loss is acceptable for the nominal mission, but we can expect the SNR performance to slightly decrease with time once on Mars due to the radiation environment. 

\begin{center}
\begin{figure}[ht]
\includegraphics[width=0.8\textwidth]{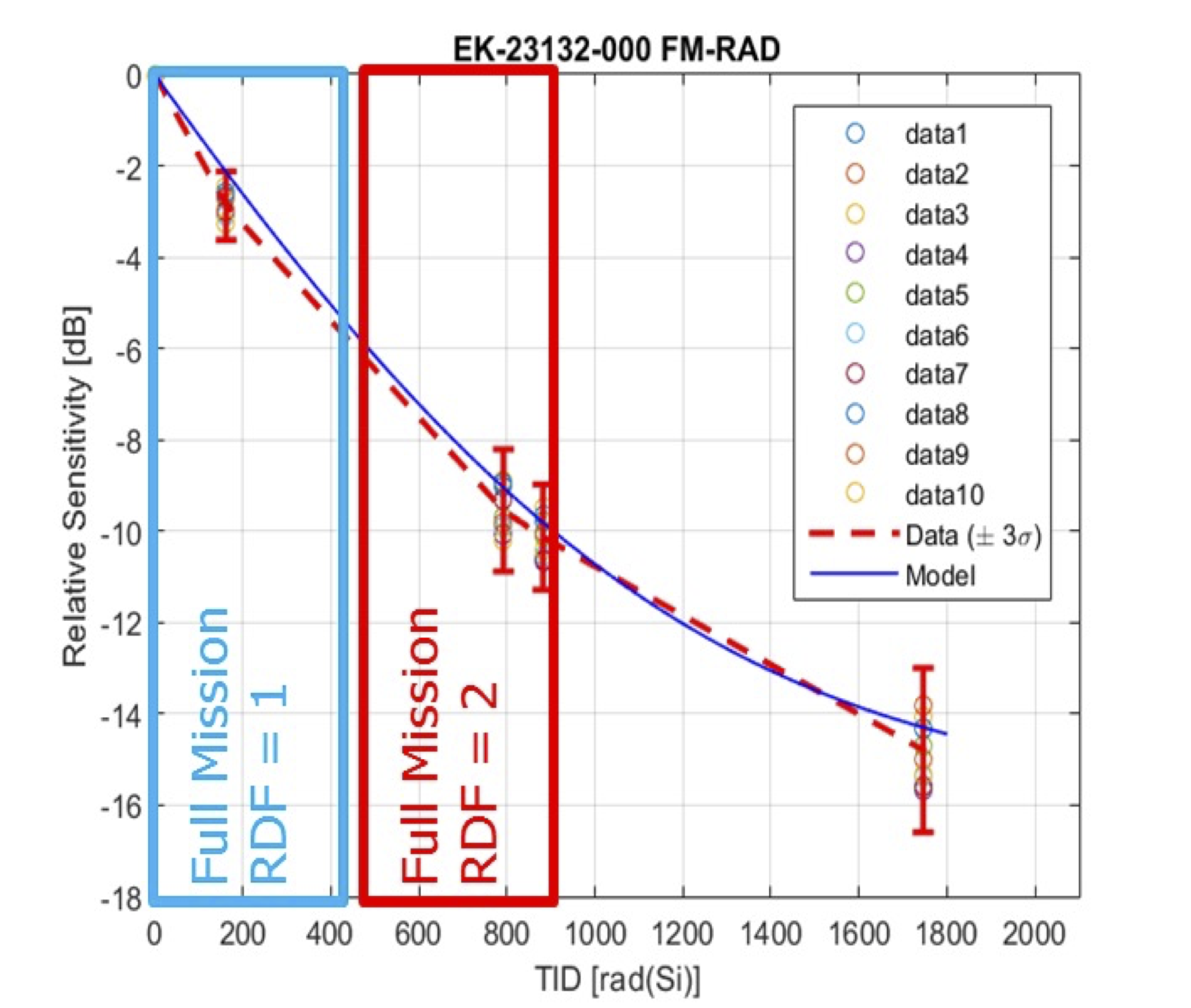}
\caption{Relative sensitivity of the EK-23132 as a function of the total ionizing dose (TID) measured during the radiation susceptibility assessment campaign)}
\label{fig:MIC-RAD}
\end{figure}
\end{center}

A group of ten microphone parts was dedicated to the total ionizing dose degradation analysis. As the microphone membrane is made of a permanently charged polymer, it is sensitive to radiation degradation. This degradation is probably due to the release of charge carriers in the medium during energy deposition. The radiation test was performed with five components powered to their nominal mission voltage of 3.3 V and five other components not powered. Components were irradiated with gamma rays from a Co60 source, and removed for testing at different Total Ionizing Dose (TID) values. The sensitivity comparison was established by comparing the components response to an acoustic sweep signal with respect to the response of a component not exposed to radiations. The excitation signal was generated by a speaker inside the test bench. The results of the relative sensitivity decay as a function of TID is presented in Figure \ref{fig:MIC-RAD}. The full mission equivalent dose is presented for radiation design factors (RDF) of 1 and 2. According to the JPL M2020 rover team analyses, the microphone will undergo most of the dose exposure during the cruise to Mars, to reach a worst case end-of-life loss of sensitivity of -10 dB after 2 years, which is still compliant with the performance requirements.

\subsection{Microphone directivity }

\begin{figure}[ht]
\center
\includegraphics[width=0.8\textwidth]{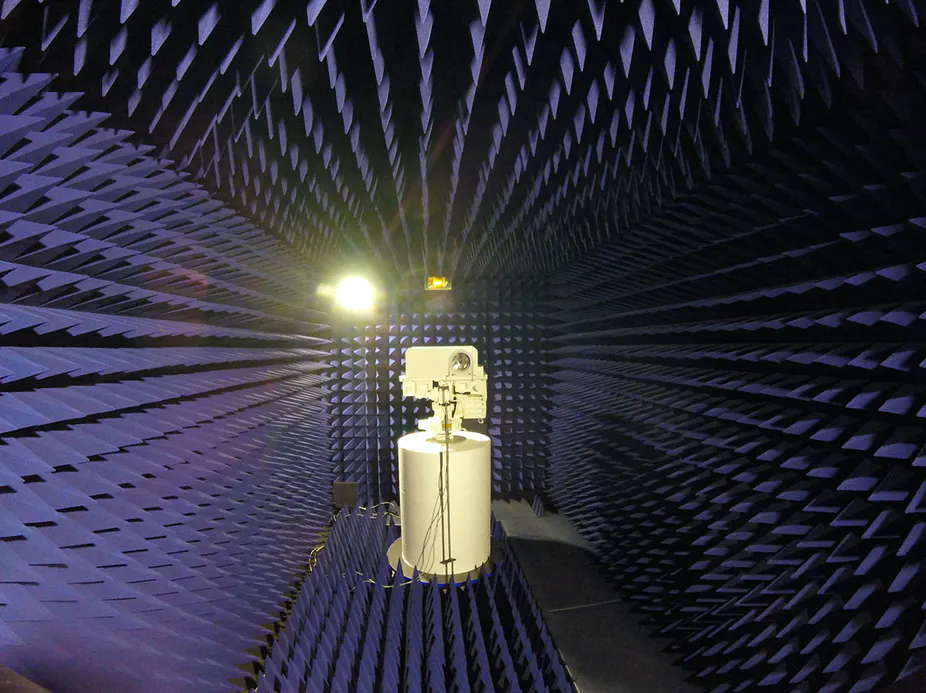}
\caption{Picture of the SuperCam Mast-Unit mock-up including the microphone in the anechoic chamber at ISAE-SUPAERO/DEOS Department}
\label{fig:MIC-Anechoic}
\end{figure}

In order to assess the directivity of the microphone measurement, we have also studied the influence of the Mast Unit orientation on the microphone recording \citep{chide2020premier}. This measurement has been done in an anechoic chamber at ISAE-SUPAERO/DEOS Department (see picture of the setup in Figure \ref{fig:MIC-Anechoic}).

\begin{figure}[ht]
\includegraphics[width=0.45\textwidth]{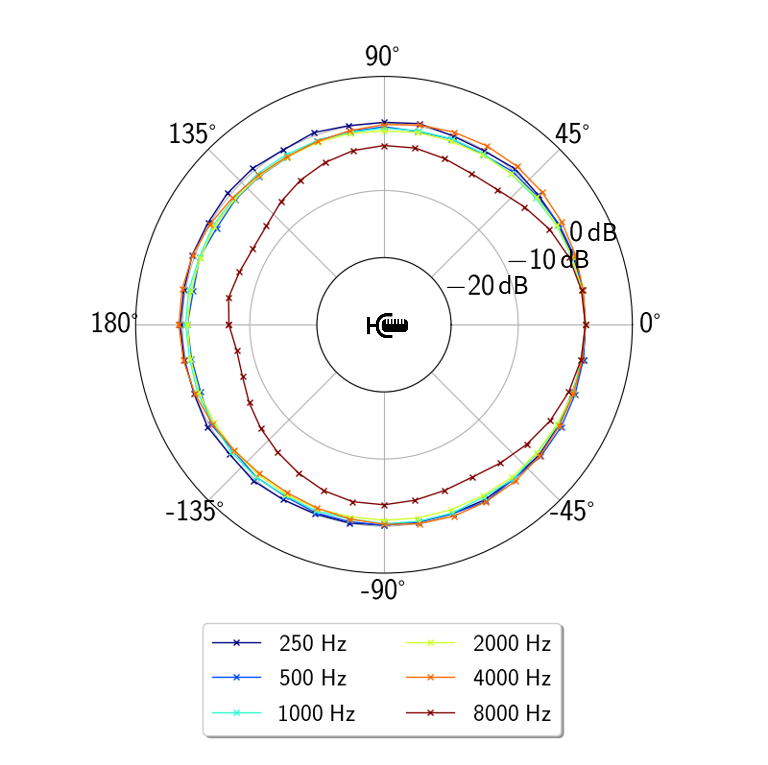}
\includegraphics[width=0.45\textwidth]{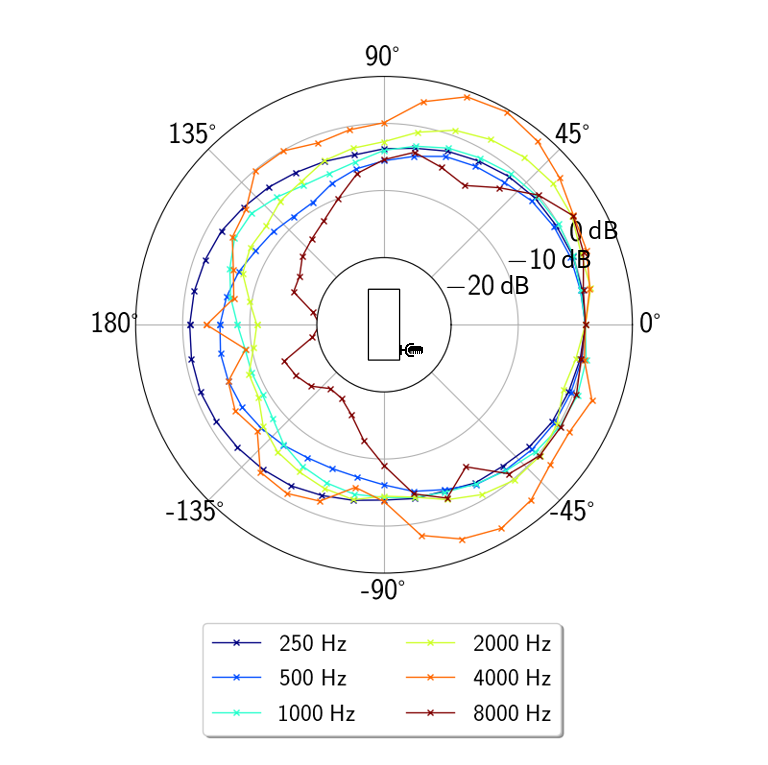}
\caption{Microphone Directivity. (Left) Directivity of the microphone Only. (Right) Directivity of the microphone implemented on the Mast Unit. Figures from \citep{chide2020premier} }.
\label{fig:fig_Directivity_MIC}
\end{figure}

The directional sensitivity of the microphone has been studied, as well as the impact of the Mast Unit on this sensitivity. Full results are described in \citep{chide2020premier} and are summarized in Figure \ref{fig:fig_Directivity_MIC}.

On the left of Figure \ref{fig:fig_Directivity_MIC}, we see that the microphone is omni-directional as specified in the microphone part datasheet. However, when the microphone is accommodated on the SuperCam Mast Unit, there is a directionality obviously linked to the presence of the Mast Unit. In the forwards-facing direction, the sensitivity remains good.

\subsection{End-to-end noise characterization and signal validation}

\begin{figure}[ht]
\includegraphics[width=0.8\textwidth]{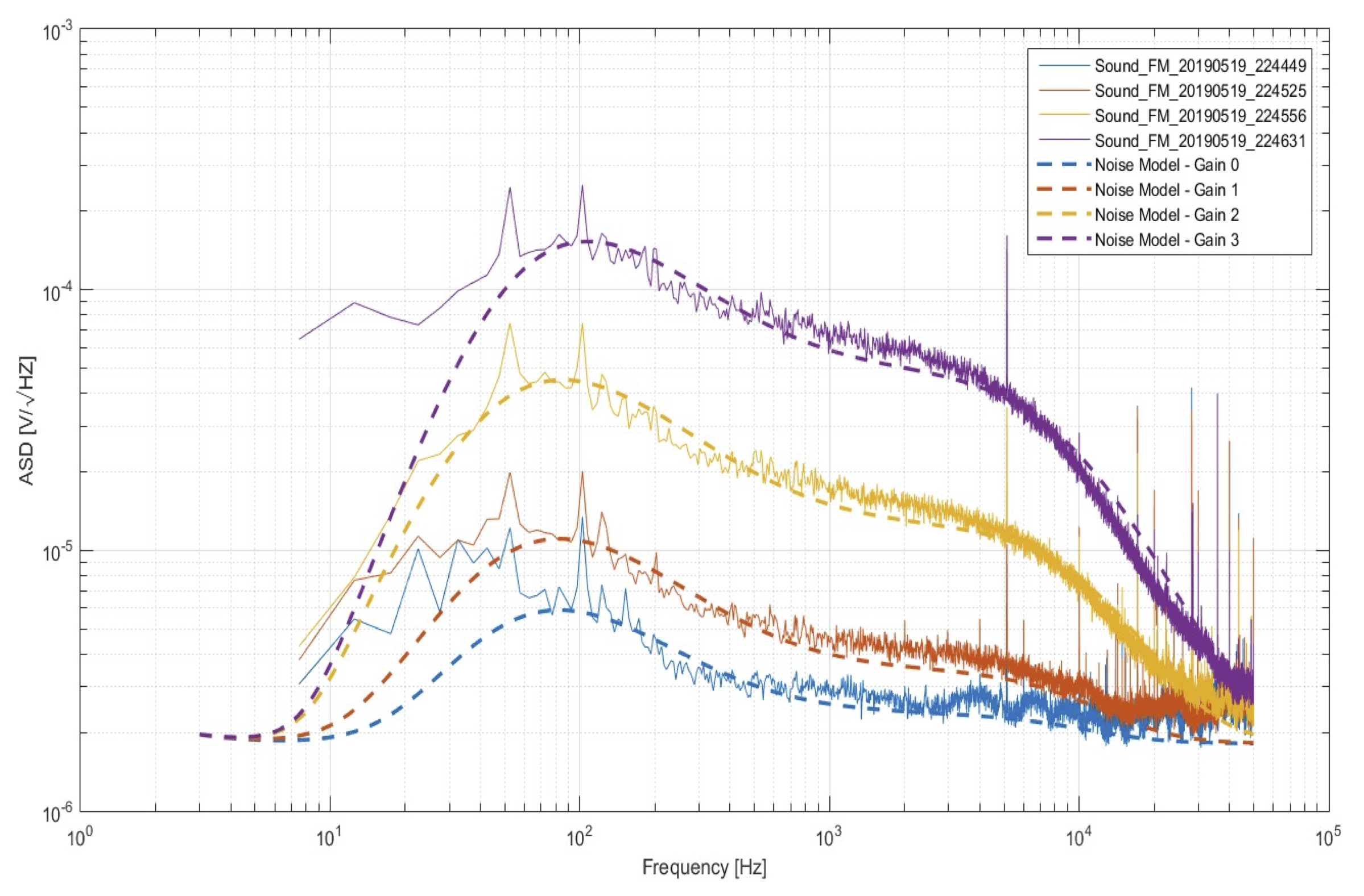}
\caption{Comparison between the noise amplitude spectral density models (noisy curve) and measurements (dashed line) of the microphone and its electronics integrated on to the SuperCam Mast Unit for all gains. The recording was performed at LESIA, in a thermal vacuum chamber under 6 mBar of N2.}
\label{fig:FEE-noise-STT}
\end{figure}

Since the clean rooms and thermal vacuum chamber are generally noisy environments in terms of acoustic measurement, the opportunities to get the complete system noise characterization are rare or non-existing. However, it happened to be possible in the LESIA facilities during the Infra-Red Spectroscope (IRS) qualification and characterization campaign. Figure \ref{fig:FEE-noise-STT} presents the comparison of the noise measurements of the microphone, integrated to the SuperCam Mast Unit, with the theoretical models for all gains, under 6 mbar of nitrogen. The model is a combination of the FEE noise model, the ADC quantification error model, and a microphone component noise model fitted with the data acquired during in the previous Martian wind tunnel test campaigns. The close match between the model and data validates the complete noise model of the SuperCam Microphone, which will be used to distinguish the intrinsic noise from the real acoustic signals in further scientific analyses.

\begin{figure}[ht]
\includegraphics[width=1\textwidth]{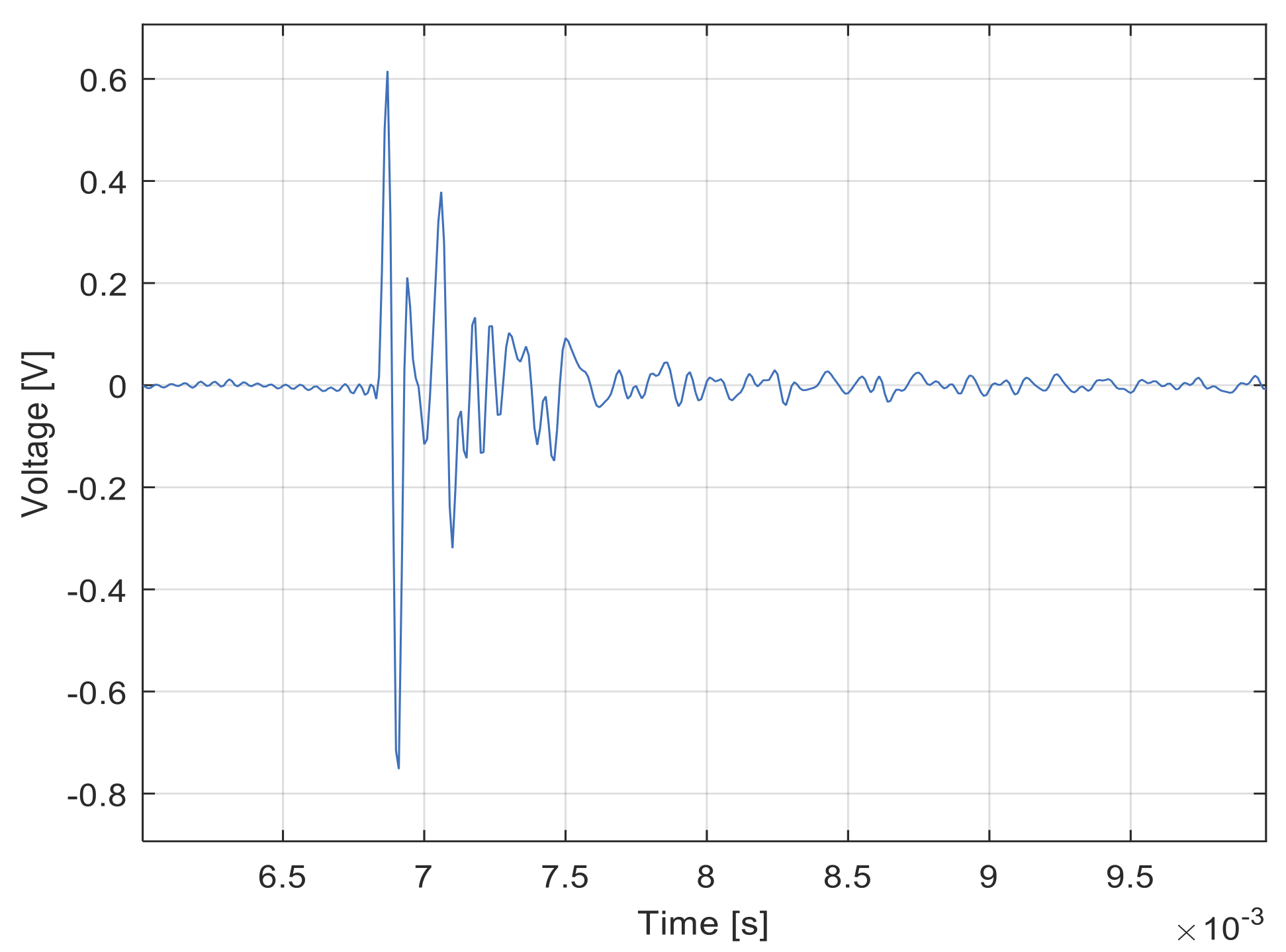}
\caption{Zoom on an acoustic waveform microphone measurement of a LIBS shot on a titanium calibration target at gain 0. The recording was performed during the system thermal test (STT) of the Perseverance rover at Jet Propulsion Laboratory (JPL), under 6 mbar of N2 at -55 °C.}
\label{fig:LIBS-shot-STT}
\end{figure}

During the system thermal test of the rover Perseverance at JPL, a raster of the LIBS laser shooting at the titanium calibration target was recorded with the microphone under a pressure of 6 mbar at -55 °C. The microphone was set to its minimum gain and successfully acquired 30 laser acoustic waveforms. A zoom on one waveform can be seen in Figure \ref{fig:LIBS-shot-STT}. The maximum amplitude is 0.7 V and the root mean square noise level is 10 mV, leading to an operation re-calibration SNR close to 60 dB, which is consequent when compared to the expected signal levels from the distant Martian rocks (20 to 30 dB). This calibration recording, and those that will be acquired on Mars, will be used to optimize the data processing pipeline currently under development, aiming at automatically extracting the relevant information from the waveforms, while removing most of the noise components.

This acquisition, just before the rover integration into the interplanetary probe, was the last signal acquired by the microphone on Earth, and it validates the complete system chain of the flight model, from the calibration target up to the rover and the ground system. 

\subsection{Mars like environment performance validation}

In order to validate the end-to-end performance of the instrument, full tests were performed using the Aarhus Wind Tunnel Simulator II \citep{holstein2014environmental} in Denmark in July 2017. The full details of these tests are reported in \citep{Murdoch2019}. The  AWTSII facility is a climatic chamber with a  wind tunnel; it has a cylindrical shape, with a 2.1 m inner diameter, and a 10 m length. The tests were performed at 6 mbar of $100\%$ $CO_2$.  A suite of environment sensors (temperature, pressure, humidity), in addition to an in-situ webcam, were used to monitor the environment.

\begin{figure}[ht]
\includegraphics[width=1\textwidth]{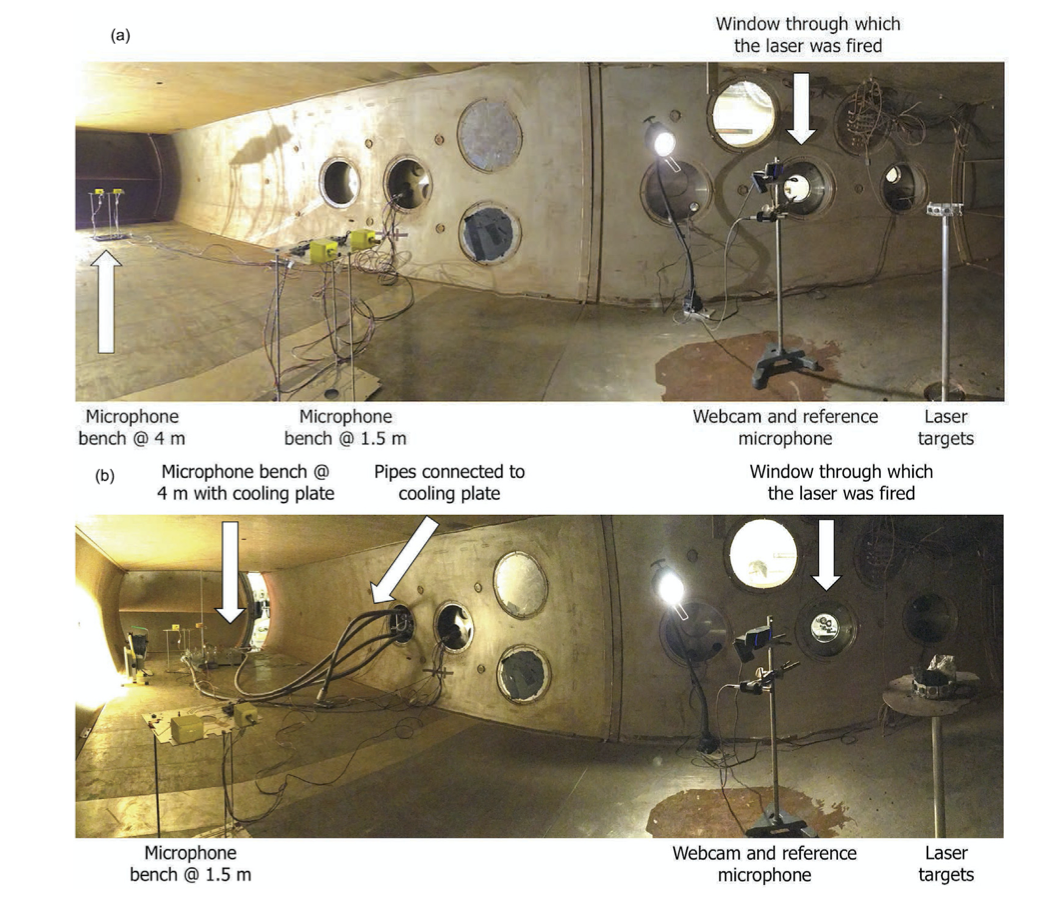}
\caption{Panoramic photos of the two test configurations at Aarhus. Above: First test configuration. Below: Second (low-temperature) test configuration. from \citep{Murdoch2019}}
\label{fig:LIBS-AarhusTestsPic-STT}
\end{figure}

These tests were the final validation that the acoustic signal of a laser LIBS blast could be recorded by the SuperCam Microphone in a martian atmosphere environment at a 4 m distance (this was the science requirement). A peak SNR of 21 dB was measured largely above the instrument SNR requirement of 10 dB. 
Thanks to this experiment we learned also two important points :
\begin{enumerate}
    \item These experiments also demonstrated that wind signal would be recorded, as it adds a high amplitude, low frequency, component to the acoustic signal. However, this wind signal can be removed with relatively simple filtering, enabling LIBS recording even in windy conditions.
    \item As a result of this, the microphone recording provide a good proxy for the wind speed, as already discussed in Section \ref{wind_speed}.
\end{enumerate}


\section{SuperCam Microphone Operations }

\subsection{Operational limitations }

They are several limitations in the use of the microphone that shall be taken into account during the ATLO process.

\begin{enumerate}

\item The maximum temperature that the microphone membrane can sustain without permanent degradation is +63 °C. Beyond that limit, the membrane polymer changes state, and the permanent electric dipole starts decreasing.

\item It has been observed that under pressure lower than 1 mBar, the microphone membrane resonance is not sufficiently damped, which creates a very strong oscillation when interacting with the electrostatic feedback force. Even though no degradation of the performances has been detected after vacuum testing, it is not recommended to power on the microphone under those conditions.

\item Membrane protection: in order to optimize its sensitivity, no filter has been implemented to protect the membrane from external objects. 

\end{enumerate}

\subsection{In Situ calibration }

In order to have the best possible performance on Mars, we have to perform in-situ calibrations, that will enable to estimate the microphone best gain tuning, the microphone sensitivity to wind as well as the performance graceful degradation along the mission.

\subsubsection{Gain calibration}

In order to determine the best gain setting strategy for atmospheric measurements, the MIC only sequence shall be played various positions of Mast Unit orientation with respect to wind. 4 sequences (gains 0, 1, 2, 3) shall be played, 100 kHz. For the LIBS recording,  LIBS continuous or LIBS pulsed mode shall be played during shots on calibration targets. 4 sequences (gains 0, 1, 2, 3) shall be played at 100 kHz. If possible, LIBS pulsed recording shall be played during every shots on calibration targets. This will help to monitor any microphone sensitivity change.

\subsubsection{Sensor cross validation}

In order to evaluate the microphone sensitivity to the wind, we propose to record a stand-alone continuous sequence with Mast rotation in azimuth together with MEDA ( 1 recording for each Mast Unit position). 

\begin{enumerate}
    
\item 	15° or 30° step in azimuth, elevation at 0°, 24 or 12 measurements in total
\item 	One 30 long microphone recording (25 kHz) per step in azimuth
\item 	Microphone measurements in parallel with MEDA to obtain  wind speed and orientation
\item 	Ideally this sequence should not be played for a wind coming from the rear of the rover, subject to too much interaction with the body of the rover and the RTG.

\end{enumerate}

\subsubsection{ Regular Calibration}

Each time a target calibration is performed, LIBS continuous recording shall be played during shots on calibration targets. 4 sequences (gains 0, 1, 2, 3) shall be played, 100 kHz. In particular, LIBS+MIC shall be played each time the SCCT Titanium is targeted: it will be used as the in-situ reference sound as acoustic LIBS signal on titanium is loud and because the ablation rate on titanium is small. Therefore the acoustic signal remains constant with the number of shots. This allows to determine the microphone gain graceful degradation as a function of time. MIC ONLY recording shall be played routinely in parallel with MEDA (one pointing only) to calibrate the microphone RMS pressure level with regard to wind speed. These measurements will also have to take into account the variation of the LIBS signal as a function of the external pressure ( which varies along the seasons).

\section{Conclusions and discussion}

The SuperCam instrument suite onboard the Mars 2020 rover includes the Mars Microphone (provided by ISAE-SUPAERO in France). The SuperCam Microphone has been the first microphone to record sounds from the surface of Mars. In order to record LIBS shock waves and atmospheric phenomena, the Mars Microphone is able to record audio signals from 100 Hz to 10 kHz on the surface of Mars, with a sensitivity sufficient to monitor a LIBS shock wave at distances of up to 4 m. It shall be an help characterize the rocks shot by SuperCam, but it also opens a new windows of atmospheric measurement on the Mars surface, providing high frequency insights on the wind and turbulence measurements. We do not have a single doubt that we have paved the way and that all Mars missions, in the coming years, will implement a microphone as a part of their atmospheric payloads.

\section{Acronyms}

The following acronyms are used in this publication :

\begin{table}[h!]
\begin{tabular}{ | m{2.5cm} | m{4cm}| m{4cm} | } 
  \hline
   \textbf{Acronym} & \textbf{Signification} & \textbf{Comment}  \\  
\hline
   \textbf{ADC}  & Analog to Digital converter  &   \\ 
  \hline
    \textbf{BU}  & Body Unit  &  SuperCam Subsystem \\ 
    \hline
    \textbf{CMOS}  & Complementary metal-oxide-semiconductor  &  \\ 
\hline
     \textbf{CNES}  & Centre National d'Etudes Spatiales  & French Space Agency \\ 
\hline

       \textbf{COTS}  & Component Off the Shelf  &   \\ 
    \hline
     \textbf{DPU}  & Digital processing Unit  & SuperCam subsystem \\ 
    \hline
    \textbf{DPA}  & Destructive Part Analysis &  \\ 
    \hline
    \textbf{DREAMS}  & Dust Characterisation, Risk Assessment, and Environment Analyser on the Martian Surface & ExoMars payload \\ 
    \hline
    
    \textbf{FEE}  & Front End Electronics  & SuperCam subsystem \\ 
    \hline
     \textbf{HK}  & House Keeping  &   \\ 
    \hline
    \textbf{IRAP}  & Institut de Recherche en Astrophysique et planétologie  & Consortium member \\ 
    \hline
     \textbf{IRS}  & Infra-Red Spectroscope & SuperCam subsystem \\ 
     \hline
     \textbf{ISAE}  & Institut Supérieur de l'aéronautique et de l'Espace & Consortium member \\ 
    \hline
    \textbf{JPL}  & Jet Propulsion laboratory  &  Lead Consortium member \\ 
    \hline
  
  \textbf{LANL}  & Los Alamos National Laboratories  &  Consortium member \\ 
    \hline
    \textbf{LIBS}  & Light Induced Breakdown Spectroscopy  &   \\ 
  \hline
  \textbf{LVPS}  & Low Voltage Power Supply  &  SuperCam subsystem \\ 
    \hline
    \textbf{LSB}  & Least Significant Bit &   \\ 
    \hline
    
   \textbf{MARDI}  & Mars Descent Imager  &  Phoenix Mission Instrument \\ 
  \hline

  \textbf{MPL}  & Mars Polar Lander  &  \\ 
  \hline
   
    \textbf{MSL}  & Mars Science Laboratory  &  \\ 
    \hline
\textbf{MU}  & Mast Unit  &   \\ 
    \hline

    \textbf{MEDA}  & Mars Environmental Dynamics Analyzer  &  Perseverance Instrument \\ 
    \hline
    \textbf{MOXIE}  & Mars Oxygen In-Situ Resource Utilization Experiment   &  Perseverance Instrument \\ 
    \hline
    \textbf{Mastcam-Z}  & Mast-Mounted Camera System  &  Perseverance Instrument \\ 
    \hline
 \textbf{PQV}  & Part Qualification Validation &  \\ 
    \hline
     \textbf{RMI}  & Remote Imager  & SuperCam subsystem \\ 
    \hline
    \textbf{OBOX}  & Optical Box  & SuperCam subsystem \\ 
    \hline

    \textbf{SET}  & Single Event Transient  &   \\ 
    \hline
    
    \textbf{SNR}  & Signal to noise ratio &   \\ 
    \hline
    \textbf{RWEB}  & Remote Warm Electronic Box  &  \\ 
  \hline

    \textbf{TID}  & Total Ionizing Dose &   \\ 
    \hline
   
\end{tabular}
\end{table}

\section{Acknowledgements}
We gratefully acknowledge funding from the French space agency (CNES), from ISAE-SUPAERO, and from Région Occitanie.


\bibliography{micro}


%

%

\end{document}